\documentclass[letterpaper,11pt]{article}
\pdfoutput=1

\usepackage{jheppub}
\usepackage{hyperref} 
\usepackage{epsfig}
\usepackage[normalem]{ulem}
\usepackage{bm}
\usepackage{bbm}
\usepackage{slashed}
\usepackage{xspace} 
\usepackage{subfigure}
\usepackage{multirow}

\newcommand{\setcaptionskip}{\setlength\baselineskip{14pt}}
\newcommand{\setmainskip}{\setlength\baselineskip{18pt}}

\usepackage{amsmath,latexsym}
\usepackage{pstricks}
\usepackage{color} 
 
\def\cN{\mathcal{N}}

\newcommand{\eq}[1]{Eq.~\eqref{eq:#1}}
\newcommand{\eqs}[2]{Eqs.~\eqref{eq:#1} and \eqref{eq:#2}} 
\renewcommand{\sec}[1]{Sec.~\ref{sec:#1}}
\newcommand{\secs}[2]{Secs.~\ref{sec:#1} and \ref{sec:#2}}

\newcommand{\fig}[1]{Fig.~\ref{fig:#1}}

\newcommand{\cL}{\mathcal{L}}
\newcommand{\cM}{\mathcal{M}}
\newcommand{\cO}{\mathcal{O}}

\newcommand{\fd}[2]{\parbox{#1}{\includegraphics[width=#1]{#2}}}
\newcommand{\Gma}[1]{\Gamma\left( #1 \right)}

\newcommand{\leftpartial}{%
  \mathrel{\vbox{\offinterlineskip\ialign{%
    \hfil##\hfil\cr
    $\scriptscriptstyle\leftarrow$\cr
    $\partial$\cr
}}}}

\newcommand{\nn}{\nonumber}
\newcommand{\mcdot}{\!\cdot\!}
\newcommand{\beq}{\begin{equation}}
\newcommand{\eeq}{\end{equation}}
\newcommand{\bea}{\begin{eqnarray}}
\newcommand{\eea}{\end{eqnarray}}

\newcommand{\Eq}[1]{Equation~\eqref{#1}}
\DeclareRobustCommand{\Sec}[1]{Sec.~\ref{#1}}

\DeclareRobustCommand{\App}[1]{App.~\ref{#1}}

\DeclareRobustCommand{\Fig}[1]{Fig.~\ref{#1}}

\DeclareRobustCommand{\Eq}[1]{Eq.~(\ref{#1})}

\newcommand\prpsq[1]{\vec{#1}^{\:2}_\perp}

\newcommand\prnth[1]{\left(#1\right)}

\newcommand{\bn}{{\bar n}}

\newcommand{\cP}{{\cal P}}
\def\bnslash{\bar n\!\!\!\slash}
\def\nslash{n\!\!\!\slash}

\renewcommand{\tabcolsep}{1ex}

\allowdisplaybreaks[1]

\begin{document}



\preprint{\vbox{\hbox{MIT--CTP 5448}}}

\title{Anomalous Dimensions from Soft Regge Constants}

\author[1]{Ian Moult,}
\author[2]{Sanjay Raman,}
\author[2]{Gregory Ridgway,}
\author[2]{Iain W. Stewart}

\affiliation[1]{Department of Physics, Yale University, New Haven, CT 06511, USA\vspace{0.5ex}}
\affiliation[2]{Center for Theoretical Physics, Massachusetts Institute of Technology, Cambridge, MA 02139, USA}

\emailAdd{ian.moult@yale.edu}
\emailAdd{sanjayra@mit.edu}
\emailAdd{gridgway@mit.edu}
\emailAdd{iains@mit.edu}

\abstract{Using an effective field theory (EFT) formalism for forward scattering, we reconsider the factorization of $2\to 2$ scattering amplitudes in the Regge limit. 
Expanding the amplitude in gauge invariant operators labelled by the number of Glauber exchanges, allows us to further factorize the standard impact factors into separate collinear and soft functions. 
The soft functions are universal, and describe radiative corrections to the Reggeized gluon states exchanged by the collinear projectiles. 
Remarkably, we find that the {\em one-loop} soft function for the single Reggeized gluon state is given to $\cO(\epsilon)$ in terms of the {\em two-loop} cusp and {\em two-loop} rapidity anomalous dimensions. 
We argue that this iterative structure follows from the simple action of crossing symmetry in the forward scattering limit, which in the EFT allows us to replace the divergent part of a soft loop by a much simpler Glauber loop. 
We use this correspondence to provide a simple calculation of the two-loop Regge trajectory using the EFT. 
We then explore its implications at higher perturbative orders, and derive the maximally matter dependent contributions to the Regge trajectory to all loop orders, i.e.~the terms $\sim \alpha_s^{k+1}n_f^k$ for any $k$, where $n_f$ is the number of massless flavors. 
These simplifications suggests that the EFT approach to the Regge limit will be helpful to explore and further understand the structure of the Regge limit.
}

\maketitle
\setmainskip

\section{Introduction}
\label{sec:intro}  

The study of limits of amplitudes provides a rare glimpse into their all loop behavior. A particularly rich limit, which has long been the subject of intense study \cite{Gell-Mann:1964aya,Mandelstam:1965zz,McCoy:1976ff,Grisaru:1974cf,Fadin:1975cb,Kuraev:1976ge,Lipatov:1976zz,Kuraev:1977fs,Balitsky:1978ic,Lipatov:1985uk,Lipatov:1995pn}, is the high energy (Regge) limit, where the Mandelstam invariants $s, t$ satisfy $|t| \ll s$. In this limit, there are non-trivial two dimensional dynamics in the transverse plane of the scattering; however, amplitudes and cross sections exhibit many simplifications when compared to general kinematic limits. Indeed, amplitudes in the Regge limit have been seen to exhibit a number of remarkable properties: They exhibit integrability \cite{Lipatov:1993yb,Faddeev:1994zg}; they can be used to extract high loop data to understand the space of functions appearing in scattering amplitudes \cite{Dixon:2012yy,DelDuca:2016lad,DelDuca:2017peo,DelDuca:2018raq,DelDuca:2018hrv}; they provide data for bootstrap approaches \cite{Dixon:2015iva,Almelid:2017qju,Caron-Huot:2019vjl}; and they can even be understood non-perturbatively at finite coupling \cite{Basso:2014pla,Bargheer:2019lic,DelDuca:2019tur}.

The simplicity of amplitudes in the Regge limit arises from their factorization into universal building blocks describing dynamics at disparate rapidities. The absence of a hard interaction further leads one to expect that the physics of the Regge limit is described entirely in terms of universal infrared quantities. In $\cN=4$ super-Yang-Mills ($\cN=4$ SYM), this occurs for generic kinematics, due to the amplitude-Wilson loop duality \cite{Drummond:2008vq,Drummond:2007au,Drummond:2008aq,Drummond:2007cf,Drummond:2007aua}. However, the more general relation between the Regge trajectory and properties of Wilson loops \cite{Korchemskaya:1996je,Korchemskaya:1994qp} shows that similar phenomena emerge in the Regge limit of QCD amplitudes.  This offers the hope that other properties, such as iterative structures, are not specific to $\cN=4$ and may also emerge in QCD. However, despite intensive study, much less is known about the general structure of amplitudes in the Regge limit beyond the case of $\cN=4$ SYM. Nevertheless, intriguing results such as \cite{DelDuca:2017pmn} suggest that much structure remains to be uncovered.

In this paper we re-examine the Regge limit of the simplest $2\to 2$ scattering amplitudes using the Glauber effective field theory (EFT)~\cite{Rothstein:2016bsq} built into the soft collinear effective theory (SCET))~\cite{Bauer:2000ew, Bauer:2000yr, Bauer:2001ct, Bauer:2001yt, Bauer:2002nz}. Taking the $gg\to gg$ process as a specific example, in the classic presentation the Regge limit of the amplitude is described up to NLL by a simple factorization~\cite{Kuraev:1976ge}:
\begin{align}\label{eq:simple_Regge}
\cM_{gg\to gg}= \left[ g_s T^c_{aa'} C^g(p_a, p_{a'})  \right]\frac{s}{t}\left[  \left( \frac{-s}{-t}   \right)^{\alpha(t)}  +\left( \frac{s}{-t}   \right)^{\alpha(t)}  \right]  \left[ g_s T^c_{bb'} C^g(p_b, p_{b'}) \right]\,. 
\end{align}
This result involves the so-called impact parameters $C^g$ as well as the Regge trajectory, $\alpha(t)$. Note that the Regge trajectory can be expressed in terms of the anomalous dimensions of Wilson lines \cite{Korchemskaya:1996je,Korchemskaya:1994qp,Bret:2011xm,DelDuca:2011ae,Falcioni:2021buo} while the $C^g$ are, in principle, arbitrary constants. Similar formulae exist for scattering involving quarks, with the same $\alpha(t)$ and different impact factors $C^q$, see for example~\cite{DelDuca:2001gu} for a detailed presentation.

In the EFT, the factorized QCD amplitude takes a different form than \Eq{eq:simple_Regge}, where it is instead expanded into gauge invariant operators labelled by the number of Glauber exchanges, each of which can be factorized into soft and collinear constants. We show that this provides a useful operator-level refactorization of the impact factors $C^i = C_s C_c^i$, which involves universal constants $C_{s}$ that are independent of the projectiles (describing radiative corrections to the Reggeized gluons or the Glauber potential) and collinear constants $C_c^i$ (which describe radiative corrections to the projectiles). The Regge trajectory $\alpha(t)$ then arises as a rapidity anomalous dimension for these soft and collinear functions~\cite{Rothstein:2016bsq}.  The factorization of the impact factors for the IR divergent contributions, which arises as a consequence of infrared factorization for scattering amplitudes applied to the Regge limit has a long history \cite{Korchemskaya:1996je,Korchemskaya:1994qp,DelDuca:2011ae,Bret:2011xm,DelDuca:2011ae,DelDuca:2013ara,Falcioni:2021dgr}. Here we use the Glauber effective theory of \cite{Rothstein:2016bsq} to also obtain a factorization for the IR finite pieces of the QCD impact factors into soft and collinear contributions.

To explore this new factorization, we first compute the one-loop soft corrections for graphs with both one and  two Glauber exchanges, denoted $S_1^{(1)}$ and $S_2^{(1)}$ respectively. Quite remarkably, we find that up to $\cO(\epsilon)$ in the dimensional regularization parameter they are completely expressed in terms of the two-loop anomalous dimensions of Wilson lines. In particular, with proper normalization, $S_1^{(1)}$ to ${\cal O}(\epsilon)$ and $S_2^{(1)}$ to ${\cal O}(\epsilon^0)$ are given by the two-loop Regge trajectory $\alpha^{(2)}$. Thus we observe a very interesting iterative structure in the soft sector. This significantly extends an earlier observation that the ${\cal O}(\epsilon^0)$ term in $S_1^{(1)}$ was exactly given by the two-loop cusp anomalous dimension~\cite{Rothstein:2016bsq}.

This is similar to the relation observed by Del Duca \cite{DelDuca:2017pmn}, who observed that the constants appearing in the one-loop quark and gluon impact factors involve the two-loop Regge trajectory. However, in his observation the impact factors are polluted by additional constants that are not related to the Regge trajectory.  We take a step forward in unraveling this iterative structure by demonstrating that the same pollution does not occur for the soft contributions to the impact factors, which are purely given by the Regge trajectory and are independent of whether we collide quarks or gluons. In our setup the contamination is associated purely with the collinear contributions which know about the identity of the projectiles.

Using the structure of the EFT, we then argue that the iterative structure for the soft constants arises naturally due to an interplay between crossing symmetry and the soft and Glauber modes in SCET. Crossing symmetry allows us to connect soft loops, which predict large logarithms, $\ln(\frac{s}{-t})$, with simpler Glauber loops, which produce factors of $i\pi$. This effectively drops the relativistic loop order by one.
Using these relations allows us to prove that the two-loop Regge trajectory must appear in $S_{1}^{(1)}$ and $S_{2}^{(1)}$, thus providing a calculation of the two-loop Regge trajectory using simpler Feynman diagrams in the EFT.

Our observations about the iterative structure of terms in the soft sector of the EFT also enable us to calculate the maximal matter-dependent terms at higher orders in $\alpha_s$, namely the terms $\alpha_s^{k+1} n_f^k$ for any $k$, where $n_f$ is the number of light flavors.
In particular we predict these maximal-$n_f$ terms in the Regge trajectory $\alpha(t)$ to all orders in $\epsilon$, providing useful information about the higher-loop behavior of the full Regge trajectory. 
We also derive 
a simple formula for the maximal-$n_f$ terms in $\alpha(t)$ which are most singular $\propto \alpha_s^{k+1}n_f^k/\epsilon^{k+1}$.
We use these results to obtain the $\alpha_s^3 n_f^2$ Regge trajectory to ${\cal O}(\epsilon^0)$ and obtain exact agreement with the recent explicit three-loop calculations of Ref.~\cite{Caola:2021rqz,Falcioni:2021dgr}.  
Finally, as a novel application of our results we give the  
$\alpha_s^4 n_f^3 \{\epsilon^{-4},\ldots, \epsilon^0\}$ and $\alpha_s^5 n_f^4 \{\epsilon^{-5},\ldots, \epsilon^0\}$ terms in the Regge trajectory. We confirm the singular terms in $\epsilon$ with the IR consistency formula of Refs.~\cite{Korchemskaya:1994qp,Korchemskaya:1996je,Bret:2011xm,DelDuca:2011ae}, while our predictions for the $\cO(\epsilon^0)$ terms are, to our knowledge, new results.

An outline of this paper is as follows. In \Sec{sec:EFT} we discuss the structure of the $2\to 2$ scattering amplitude in the Glauber EFT, emphasizing the difference between our factorization and the classic factorization into impact factors and a Regge trajectory.  We also provide perturbative results for the soft constants for one and two Glauber exchanges. 
Next, in \Sec{sec:relations} we show that basic unitarity relations have an interesting interpretation in the EFT, where they relate amplitudes with different numbers of loops of soft and Glauber modes. 
The Glauber loops are simple and effectively constrained by unitarity constraints on the forward scattering cross section.
The relation between Glauber and soft loops thus effectively enables us to ``drop'' the loop order of a given diagram by one order. Using this, we explain some of the iterative structure appearing in the Regge trajectory and soft contributions to impact factors; and we also provide an extremely simple calculation of the Regge trajectory to two loops.
In \Sec{sec:3loop_beyond} we predict the maximal-$n_f$ terms in the Regge trajectory $\alpha(t)$ to all orders in $\alpha_s$.  An overall discussion and conclusion is found in \Sec{sec:conc}.

\section{The Factorized Two-to-Two Forward Scattering Amplitude}
\label{sec:EFT}

In this section, we investigate the factorized structure of the $2\to 2$ scattering amplitude in the Glauber effective theory, which will allow us to express the scattering amplitude as a sum of gauge invariant operators. These operators have rapidity anomalous dimensions that are related to Regge trajectories. 
This relationship was demonstrated at leading logarithmic order for the gluon Regge trajectory in \cite{Rothstein:2016bsq}. 
Our focus here will be on the structure of the ``constants'' in the forward scattering amplitudes, which we define to be the terms not predicted by the renormalization group evolution (rapidity and UV divergences).
A key feature of the Glauber EFT, which will prove helpful in improving the understanding of the structure of the impact factors, is that the EFT further factorizes the impact factors into soft and collinear constants, each of which individually arise from different gauge invariant operators in the EFT. In particular, we will see that the soft constants are purely associated with the Glauber potential (and therefore the Regge trajectory), while the collinear constants are associated with the dynamics of the projectiles. This further factorization will significantly simplify a number of the observations of Del Duca~\cite{DelDuca:2017pmn}. 

We begin by discussing the factorization of the $2\to 2$ scattering amplitude in \Sec{sec:fact}. This section is not meant to provide completely detailed expressions for generic contributions; instead, it serves to illustrate the nature of the factorization into soft and collinear components. We then focus in detail on the $8_A$ color channel in \secs{soft}{collinear}. After factorizing the impact factors into collinear and soft contributions to constants, we then explicitly calculate the constants appearing in the soft functions.

\subsection{The Factorized Amplitude from the EFT}\label{sec:fact}

The Regge limit is defined by a power-counting parameter $\lambda = \sqrt{-t/s}$, and can be described by SCET~\cite{Bauer:2000ew, Bauer:2000yr, Bauer:2001ct, Bauer:2001yt, Bauer:2002nz} supplemented by operators incorporating Glauber potentials. Since there is no hard scattering process in the Regge limit, we have
\begin{align} \label{eq:SCETLagExpand}
\cL_{\text{SCET}}= \cL_{n\bn s}^{(0)} + \cL_G^{(0)}\,,
\end{align}
where 
\begin{align}
{\cal L}_{n\bn s}^{(0)}={\cal L}_n^{(0)}+{\cal L}_\bn^{(0)}+{\cal L}_s^{(0)}\,,
\end{align}
describes the factorized dynamics of soft and collinear degrees of freedom in SCET, 
and~\cite{Rothstein:2016bsq}
\begin{align}\label{eq:Glauber_Lagrangian}
\cL_G^{\text{II}(0)} 
&= e^{-ix\cdot \cP} \sum\limits_{n,\bar n} \sum\limits_{i,j=q,g}   \cO_n^{iB} \frac{1}{\cP_\perp^2} \cO_s^{BC}   \frac{1}{\cP_\perp^2} \cO_{\bar n}^{jC}   + e^{-ix\cdot \cP} \sum\limits_{n} \sum\limits_{i,j=q,g} \cO_n^{iB}   \frac{1}{\cP_\perp^2} \cO_s^{j_n B}\,,
\end{align}
is the leading power Glauber Lagrangian. Here $\cO_n^{iB}$, $\cO_{\bar n}^{jC}$ denote particular collinear operators, while $\cO_s^{BC} $ and $\cO_s^{j_n B}$ denote soft operators.  The sums over $i,j$ include quark and gluon operators. Their explicit form is not required for the current discussion, but can be found in~\cite{Rothstein:2016bsq}. The important feature of the Glauber Lagrangian is that it couples the $n$ and $\bar n$ degrees of freedom (undergoing the scattering in the Regge limit) together with internal soft degrees of freedom.

As shown in~\cite{Rothstein:2016bsq} one can derive factorized expressions for the $2\to 2$ scattering amplitude to any logarithmic order by expanding it as a sum over the number of insertions of the Glauber potential. In perturbation theory, this is a meaningful expansion, since the Glauber potential loops do not give logarithms, and so the addition of extra Glauber potential operators reduces the logarithmic order of a perturbative correction. Once the number of Glauber operators is fixed, the soft and collinear operators can be factorized to all orders in $\alpha_s$.

\begin{figure}
\begin{center}
\includegraphics[width=0.5\columnwidth]{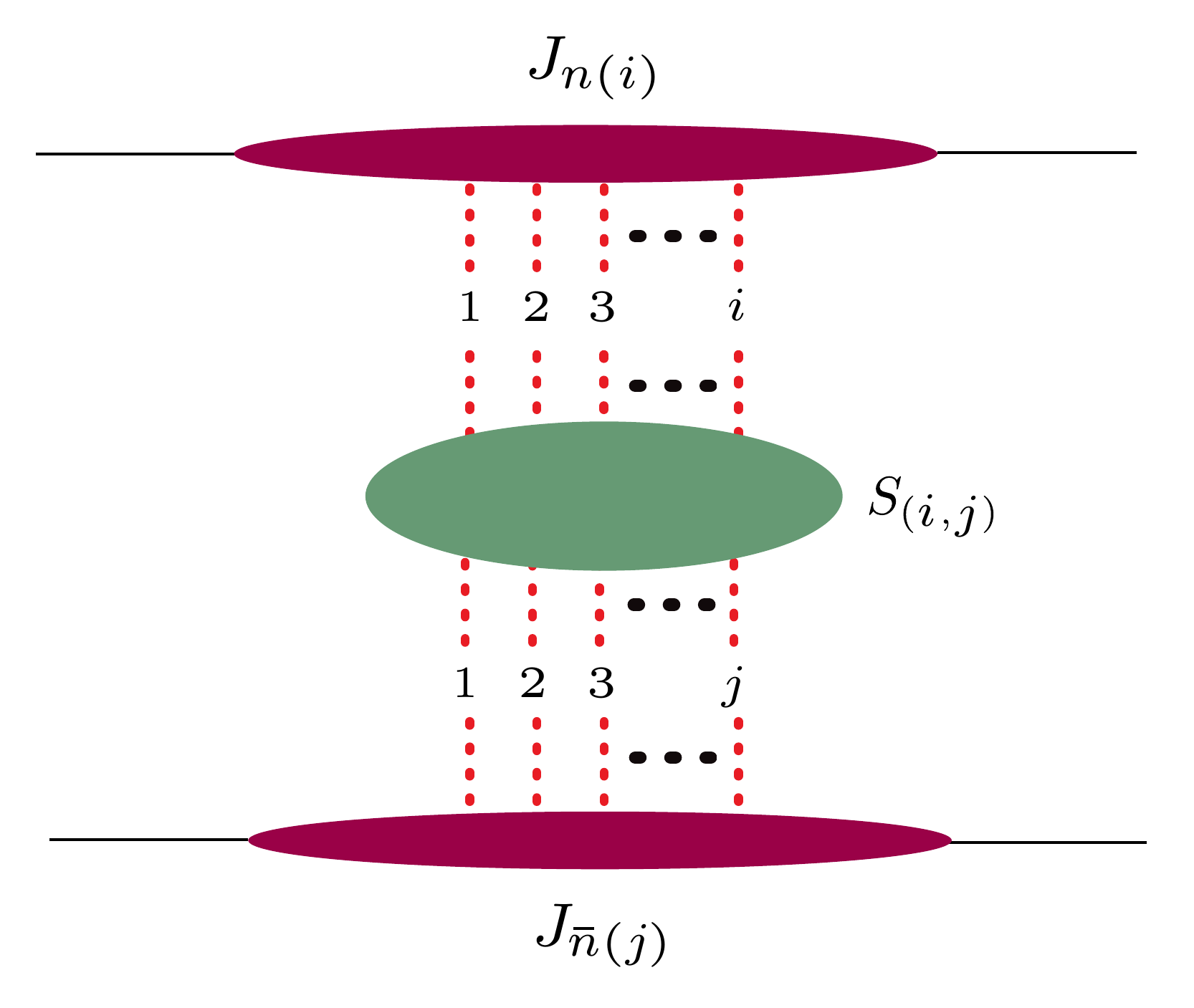}
\end{center}
	\vspace{-0.4cm}
	\caption{\setcaptionskip Factorized structure of the $2\to2$ forward scattering amplitude in the Glauber EFT. The amplitude is expressed as a sum of the number of Glauber exchanges. For a fixed number of Glaubers,  the collinear factors describe the interaction of the projectiles with the Glaubers, while the soft factor describes radiative corrections to the Glauber potential. }
	\label{fig:fact_diagram}
	\setmainskip
\end{figure}

More precisely, starting with the time evolution operator in the EFT
\begin{align}
	U(a,b;T ) & = \lim_{T\to \infty (1-i0)} \int \big[{\cal D}\phi\big] \exp\bigg[ i \int_{-T}^{T}\!\! d^4x\: \big( {\cal L}_{n\bn s}^{(0)}(x) + {\cal L}_G^{\rm II(0)}(x)\big) \bigg] \,,
\end{align}
one can expand in the number of Glauber potential insertions attaching to the $n$ and $\bn$ projectiles, given by $i$ and $j$ respectively, so that
\begin{align}
 \exp\bigg[ i \int_{-T}^{T}\!\! d^4x\: \big( {\cal L}_G^{\rm II(0)}(x)\big) \bigg]= 1 +\sum\limits_{i=1}^\infty \sum\limits_{j=1}^\infty U_{(i,j)}.
\end{align}
For any number of Glauber potential insertions, one can then factorize the soft and collinear operators to give a factorized expression for the amplitude that can be schematically written as
\begin{align}\label{eq:general}
	i\mathcal{M}_{2\to2} 
 &= \sum_{i,j=1}^{\infty} J_{n(i)}^\kappa(\nu) \otimes S_{(i,j)}(\nu) \otimes J_{\bn(j)}^{\kappa'} (\nu)
 \nn\\
  &= \sum_{i,j=1}^{\infty} J_{(i)}^\kappa(\nu) \otimes S_{(i,j)}(\nu) \otimes J_{(j)}^{\kappa'} (\nu)
  \,,
\end{align} 
where $J_{n(i)}^\kappa$ is a matrix element of $n$-collinear fields for the projectile $\kappa$ to scatter with $i$ Glauber operators, $J_{\bn(j)}^{\kappa'}$ is the analog for the other projectile with $\bn$-collinear fields, and $S_{(i,j)}$ encodes all the soft dynamics for this configuration.
Here the $\otimes$ denote integrals over transverse momenta arguments in the various $J_n$, $J_\bn$, $S$ factors that are suppressed for simplicity, $\kappa, \kappa'= q,g$ for quark or gluon projectiles, and the dimension-1 parameter $\nu$ is known as the rapidity renormalization scale. In the second line we dropped the $n$,$\bn$ subscripts, since the symmetry under $n\leftrightarrow \bn$ implies $J_{n(i)}^\kappa=J_{\bn(i)}^\kappa\equiv J_{(i)}^\kappa$.
This factorization is illustrated in \Fig{fig:fact_diagram}. The rapidity renormalization group evolution in $\nu$ of the different operators enables a resummation of logarithms of $t/s$, and gives rise to Reggeization.

While \Eq{eq:general} is completely general, in this paper we will be interested in understanding its implications for the Regge trajectory to two loop order for the antisymmetric-octet ($8_A$) color representation, which is suitable for resummation at next-to-leading logarithmic (NLL) order. By carrying out a t-channel color decomposition we can project \eq{general} onto the $8_A$ channel. For this analysis we can restrict to  exchanges with $i,j\leq 2$ Glaubers. In \sec{collinear} we demonstrate that the collinear projectile factors are universal for one and two Glaubers,  so that
\begin{align} \label{eq:J1isJ2}
  J_{(1)}^{\kappa(8_A)}=J_{(2)}^{\kappa(8_A)}\equiv J^{\kappa(8_A)}
  \,.  
\end{align}
For generic scattering projectiles $\kappa,\kappa'$, we therefore have%
\footnote{In principle there is also a contribution $S^{(8_A)}_{(2,1)}+S^{(8_A)}_{(1,2)}$ to consider which starts at the same order in $\alpha_s$ as $S^{(8_A)}_{(2,2)}$. Both the real and imaginary part of these contributions vanish in the EFT~\cite{Sanjayfuture}. However, any non-zero real part does not actually matter for the analysis here, as it could simply be absorbed into a perturbative correction to $S^{(8_A)}_{(1,1)}$.}
\begin{align} \label{eq:factor8A}
	i \mathcal{M}^{(8_A)}_{2\to2} \Big|_\text{NLL} = \left(\frac{s}{-t}\right)^{\alpha(t)}        J^{\kappa(8_A)}(\nu = \sqrt{s})   J^{\kappa'(8_A)}(\nu = \sqrt{s})  \left[ S^{(8_A)}_{(1,1)}(\nu = \sqrt{-t})+ S^{(8_A)}_{(2,2)}(\nu = \sqrt{-t}) \right] \, .
\end{align}
Here the collinear and soft functions are evaluated at the scales $\nu$ for which they do not contain large logarithms of $s/t$.  The $\big(\frac{s}{-t}\big)^{\alpha(t)}$ prefactor comes from the renormalization group evolution. The anomalous dimension here is determined by the gluon Regge trajectory $\alpha(t)$.  For the particular case of the $8_A$ color channel with two-Glauber exchange, the $\otimes$ factors are no longer present to this order. This is a specific feature of the effective field theory for forward scattering, and highlights the difference between ``Glauber gluons" and ``Reggeons". To understand this, note that for Reggeons, there is no $8_A$ two-Reggeon state. This is because it is related by crossing symmetry to the single Reggeon exchange, and can be absorbed by shifting the argument of the logarithm (see e.g.~\cite{Caron-Huot:2017fxr} for a detailed discussion). It can therefore be viewed as dressing the single Reggeon exchange, which is a pure pole, with $i\pi$ terms. In the effective theory, there is a two-Glauber $8_A$ state. However, 
it leads to renormalization group evolution which is fixed by the same pure pole solution as for single Glauber exchange, which is necessary in order for it
to be consistent with the Reggeon picture. This allows the convolution to be eliminated at this order for this specific color channel. Note that here, $S^{(8_A)}_{(1,1)}(\nu = \sqrt{-t})$ involves a single Glauber exchange and is purely real, while $S^{(8_A)}_{(2,2)}(\nu = \sqrt{-t}) $ involves two Glauber exchanges with a single Glauber potential loop and is purely imaginary.  As compared with the standard Regge factorization formula \Eq{eq:simple_Regge}, we have factorized the impact factors into contributions from $J$ and $S$, each of which have operator definitions in the EFT. While this may seem to be minor, we will see that it leads to considerable insight into the iterative structure of these functions.

Having presented a new factorized expression for the $2\to 2$ scattering amplitude in the EFT, it is now interesting to understand the structure of the radiative corrections to the factorized components. In \Sec{sec:soft} we compute the one loop soft corrections to the Glauber potential, which we will ultimately show are closely related with the Regge trajectory itself. The nature of the collinear functions is briefly discussed in \Sec{sec:collinear}.

\subsection{One Loop Soft Corrections to the Glauber Potential}\label{sec:soft}

In this section we consider the one-loop soft corrections to the $1\to 1$ and $2\to 2$ Glauber potentials.

\subsubsection{Notation and Perturbative Constants}

Our results for the soft corrections to the Glauber potential simplify considerably when expressed in terms of a variety of commonly used constants. Here we briefly summarize these constants so that our later presentation of the results can be made as compact as possible. 

Since Glauber integrals reduce to effective two dimensional integrals in the transverse space, we will often encounter the two-dimensional euclidean Glauber box integral, expressed in terms of a bubble integral
\begin{align}
\int \frac{d^{2-2\epsilon}\vec k_\perp}{(2\pi)^{2-2\epsilon}}
  \frac{1}{\big(\vec k_\perp^{\,2}\big)^\alpha 
   \big[\bigl(\vec q_\perp \!+\vec k_\perp\!\bigr)^2\, \big]^\beta}
   =\frac{B(\alpha,\beta)}{(4\pi)^{1-\epsilon}} 
   \big(\vec q_\perp^{\,2}\big)^{1-\epsilon-\alpha-\beta}, 
\end{align}
where 
\begin{align} \label{eq:Bab}
    B(\alpha, \beta) \equiv \frac{\Gma{1-\alpha-\epsilon} \Gma{1-\beta-\epsilon} \Gamma(\alpha+\beta-1+\epsilon)}{
    \Gma{\alpha} \Gma{\beta} \Gma{2-\alpha-\beta -2\epsilon}
    } . 
\end{align}
We will also find it convenient to use a modified coupling $\tilde \alpha_s$, which is often referred to as the high energy coupling, see e.g.~\cite{DelDuca:2017pmn}. It is defined in terms of the standard bare coupling $\alpha_s$ or renormalized coupling $\alpha_s(\mu)$ as follows: 
\begin{align}
    \tilde{\alpha}_s &=-\frac{\epsilon}{2}B(1,1)  \left( \frac{\prpsq{q}}{4\pi} \right)^{-\epsilon} \alpha_s 
  = -\frac{\epsilon}{2}B(1,1)  Z_\alpha \left(\frac{\mu^2 e^{\gamma_E}}{\prpsq{q}} \right)^\epsilon 
   \alpha_s(\mu)
 \nn\\
  &= Z_\alpha \left(\frac{\mu^2 e^{\gamma_E}}{\prpsq{q}} \right)^\epsilon \frac{\Gma{1-\epsilon}^2 \Gma{1+\epsilon}}{\Gma{1-2\epsilon}} \alpha_s(\mu) \nn\\
    & = Z_\alpha \left(\frac{\mu^2 }{\prpsq{q}} \right)^\epsilon 
      \Big( 1 - \frac{\pi^2}{12} \epsilon^2 - \frac{7\zeta_3}{3} \epsilon^3 +\ldots
      \Big) \alpha_s(\mu) \,,
\end{align}
where $Z_\alpha$ is the coupling renormalization factor in $\overline{\rm MS}$.
The coupling $\tilde\alpha_s$ absorbs common $\epsilon$ dependence from the Glauber box integral.
 We will also find it convenient to normalize some of our results to the following ratio of bubble integrals
\begin{align} \label{eq:Aeps}
A(\epsilon)=\frac{3}{4}\frac{B(1,1)}{B(1,1+\epsilon)}=1 + 6\zeta_3 \epsilon^3+\cdots\,.
\end{align}
At one loop, where the leading divergences are $1/\epsilon^2$, this only modifies the $\cO(\epsilon)$ terms. 

Using the modified coupling, we will be able to write our one loop results entirely in terms of constants appearing in the two-loop Regge trajectory. Expanding the Regge trajectory perturbatively in terms of the high energy coupling as 
\begin{align} \label{eq:alphaexpn}
\alpha(t)=\sum_{L=1}^\infty \left( \frac{\tilde \alpha_s}{4\pi} \right)^L \alpha^{(L)}(\epsilon)\,,
\end{align}
we have the following expressions up to two-loop order \cite{Fadin:1995xg,Fadin:1996tb,Fadin:1995km,Blumlein:1998ib}:
\begin{align} \label{eq:gregge}
\alpha^{(1)}(\epsilon)&=\frac{2C_A}{\epsilon}\,, \\
\alpha^{(2)}(\epsilon)&=\frac{\beta_0}{\epsilon^2}C_A+2 \frac{ \Gamma_\text{cusp}^{(1)}}{\epsilon}C_A + \Gamma_R^{(1)}
 \,.
\end{align}
We also expand the $\beta$-function and cusp anomalous dimension as
\begin{align} \label{eq:betafunction}
\beta(\alpha_s) =
- 2 \alpha_s \sum_{n=0}^\infty \beta_n\Bigl(\frac{\alpha_s}{4\pi}\Bigr)^{n+1}
\,,\qquad
\Gamma_{\rm cusp}(\alpha_s) =
\sum_{n=0}^\infty \Gamma^{(n)}_{\rm cusp} \Bigl(\frac{\alpha_s}{\pi}\Bigr)^{n+1}
\,,\end{align}
where the one and two-loop cusp anomalous dimensions are~ \cite{Korchemsky:1987wg}
\begin{align}
    \Gamma_\text{cusp}^{(0)} &= 2 \,, 
  &  
  \Gamma_\text{cusp}^{(1)} &= -\frac{5n_f}9 + C_A \left(\frac{67}{18}-\frac{\pi^2}{6}\right)\,,
\end{align}
$\beta_0 = (11 C_A - 4 T_F n_f)/3$ is the one-loop $\beta$-function, $C_A$ and $C_F$ are the quadratic Casimirs for the adjoint and fundamental representations, respectively, $n_f$ is the number of quark flavors, and $T_F=1/2$. Finally $ \Gamma_R^{(1)}$ is the ${\cal O}(\epsilon^0)$ term in the two-loop Regge trajectory and is given by  
\begin{align}
    \Gamma_R^{(1)} &= -\frac{56 C_A n_f}{27} + C_A^2 \left( \frac{404}{27} - 2 \zeta_3 \right) \,.
\end{align}
In a generic gauge theory, the relation between $\Gamma_R^{(1)}$ and other anomalous dimensions is not known. In $\cN=4$ SYM, $\Gamma_R^{(1)}$ is equivalent to the collinear anomalous dimension \cite{Drummond:2007aua}. In QCD an intriguing relation to certain anomalous dimensions of cusped Wilson lines \cite{Erdogan:2011yc} has been observed to two-loops, but it is unclear if this persists to higher order. For a recent discussion of relations between anomalous dimensions, see \cite{Falcioni:2019nxk}. 

\subsubsection{Single Glauber Potential}

We begin by considering the radiative corrections to the single Glauber potential at one-loop. These calculations were first performed in \cite{Rothstein:2016bsq} using a mass regulator. Here we use dimensional regularization for infrared singularities; since the setup for the calculations is analogous to \cite{Rothstein:2016bsq}, we provide few details. We will also extend the calculation to determine the $\cO(\epsilon)$ terms, inspired by the observations of \cite{DelDuca:2017pmn}.

We expand the single Glauber potential perturbatively as
\begin{align} \label{eq:S1expn}
S_{(1,1)}^{(8_A)} \equiv S_1
 = S_{1}^{(0)}  \bigg[ 1 + \left( \frac{\tilde \alpha_s}{4\pi} \right) S_{1}^{(1)}+\left( \frac{\tilde \alpha_s}{4\pi} \right)^2 S_{1}^{(2)} +\cdots \bigg]  \,,
\end{align}
where the tree level Glauber exchange potential is given by
\begin{align} \label{eq:S10}
    S_{1}^{(0)} &= \left(\fd{2cm}{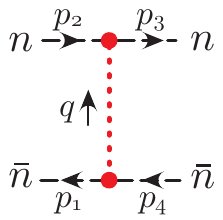}\right)^{(8_A)} 
     = -\frac{8 \pi i \alpha_s }{\prpsq{q}} 
     = \frac{8 \pi i \alpha_s }{t} \,.
\end{align}
This will be used to normalize our results. In $S_1^{(i)}$ the subscript 1  denotes that this is the single Glauber exchange, and the superscript $(i)$ denotes the number of soft loops. 
More generally $S_j^{(i)}$ will denote coefficients of the ${\cal O}(\alpha_s^{i+j})$ term from the sum of graphs with $j$ Glauber exchanges and $i$ soft loops. 
Below we will always use the notation $t=-\vec q_\perp^{\,2}$. 
Note that we use a dashed line for the projectiles in Feynman diagrams, which will however always be transverse gluons for our calculations. We also always strip off a common polarization dependent prefactor in our Feynman graphs, which is $(s\, \vec\epsilon_2^\perp\cdot \vec\epsilon_3^\perp \vec\epsilon_1^\perp\cdot \vec\epsilon_4^\perp)$, where $s= n\cdot p_1 \bn\cdot p_2$ to leading order in the $|t|\ll s$ expansion.
Finally, \eq{S10} defines our normalization convention for projection onto $(8_A)$ as this is the only color channel for this diagram.

At one-loop there are two corrections to the Glauber potential, involving either a soft quark loop, or a soft gluon loop. Although these look like pure vacuum polarization graphs, the gluon loop graph involves both vacuum polarization and Wilson line contributions at this order. Furthermore, since the result is formulated in terms of gauge invariant operators in the EFT, ghost loops are not required at this order, and the gluon loop also has rapidity divergences from the presence of Wilson lines in the soft gluon operator. We use the rapidity regulator $\eta$, and expand results in the limit $\eta\to 0$ to separate the rapidity divergent and constant terms, see~\cite{Chiu:2012ir,Rothstein:2016bsq}. Performing the calculations, we find that
\begin{align} \label{eq:Sqeye}
    \left(\fd{2cm}{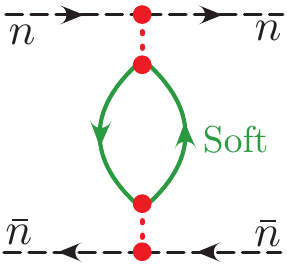}\right)^{(8_A)} &= 
  -2 n_f T_F S_1^{(0)}  \frac{\alpha_s(\mu)}{\pi} 
    \left(\frac{\mu^2 e^{\gamma_E}}{-t} \right)^\epsilon 
    \frac{\Gma{2-\epsilon}^2}{\Gma{4-2\epsilon}}  \Gma{\epsilon} 
    \nn \\
    &= S_{1}^{(0)} n_f T_F A(\epsilon) \prnth{\frac{\tilde{\alpha}_s}{4\pi}}  \Big( 
    -\frac{4}{3\epsilon} - \frac{20}9 - \frac{112}{27} \epsilon
      + \mathcal{O}(\epsilon^2) \Big)\,,
\end{align}
and
\begin{align} \label{eq:Sgeye}
    \left(\fd{2cm}{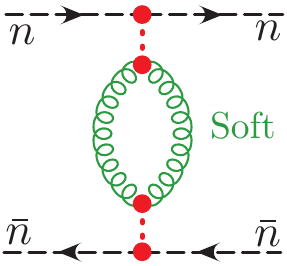}\right)^{(8_A)} &= 
    -\frac{1}{2} \frac{\alpha_s(\mu)}{\pi} C_A S_{1}^{(0)}
    \left(\frac{\mu^2 e^{\gamma_E}}{-t} \right)^\epsilon
    \bigg\{
    \frac{\Gma{2-\epsilon}^2}{\Gma{4-2\epsilon}}  \Gma{\epsilon} 
    - 2\frac{\Gma{1-\epsilon}^2}{\Gma{2-2\epsilon}}  \Gma{\epsilon}
    \nn \\
    & \;\;\;\;\;\; +\frac{\Gma{\frac{\eta}2}\Gma{\frac{1-\eta}2} \Gma{1+\epsilon + \frac\eta2} \Gma{-\epsilon-\frac\eta2 } }{\Gma{1+\frac\eta2} \Gma{\frac12-\epsilon-\frac\eta2}} 
    4^\epsilon \left( \frac{\nu^2}{-t}\right)^{\frac\eta2}
    \bigg\}
 \,,
\end{align}
which expanding for $\eta\to 0$ and setting $\nu=\sqrt{-t}$ gives the desired constant terms
\begin{align} \label{eq:vacpolren}
   \left(\fd{2cm}{figsSC/Glaub_softvac_gluons_ext}
    \right)^{(8_A)}_{\stackrel{\!\!\!\!\mathcal{O}(\eta^0)}{\nu=\sqrt{-t}} }
      &=  \prnth{\frac{\tilde{\alpha}_s}{4\pi}} C_A S_{1}^{(0)} A(\epsilon)  \biggl[ 
    -  \frac2{\epsilon^2} + \frac{11}{3 \epsilon} - \frac{\pi^2}{3} + \frac{67}{9}
    +\left(\frac{404}{27} - 2 \zeta_3 \right) \epsilon 
     + \mathcal{O}(\epsilon^2)  \biggr] .
\end{align}
Here the $1/\epsilon^k$ terms are IR divergences, and the soft constant includes both these and infrared finite terms.
The SCET Glauber operator with two gluons generates a number of terms in \eq{Sgeye}, including standard vacuum polarization contributions as well as terms with eikonal denominators that lead to rapidity divergences, see Ref.~\cite{Rothstein:2016bsq} for details.
The subscripts on the graph in \eq{vacpolren} remind us that the soft constants are defined as renormalized quantities 
in a scheme that minimally subtracts the $1/\eta$ rapidity divergences.
The subscript $\mathcal{O}(\eta^0)$ indicates that in the definition of the soft constants we cancel the $\eta$ divergent terms against the $\eta$ divergent terms from the collinear sector. 
Additionally, we take the rapidity scale $\nu= \sqrt{-t}$ to eliminate large logarithms in the soft function, as dictated by \eq{factor8A}.

Notice also that we have kept terms up to $\mathcal{O}(\epsilon)$. 
At two-loops we will see that these terms will iterate such that they multiply the $\frac1\epsilon$ divergent $\alpha^{(1)}$ in \eq{gregge}, and hence  contribute to the constant piece of the two-loop Regge trajectory.

Summing the above two soft graphs then gives us the 1-loop, 1-Glauber soft constant
\begin{align}
    & \hspace{-2cm}
    \left(\fd{2cm}{figsSC/Glaub_softvac_quark_ext}\right)^{(8_A)} \; + \; \left(\fd{2cm}{figsSC/Glaub_softvac_gluons_ext}\right)^{(8_A)}_{\stackrel{\!\!\!\!\mathcal{O}(\eta^0)}{\nu=\sqrt{-t}} }
     \nn \\[5pt]
    &=  S_1^{(0)} \frac{\tilde{\alpha}_s}{4\pi} A(\epsilon)  \biggl[
    - \frac{2 C_A}{\epsilon^2} 
    +\frac{(11C_A - 2n_f)}{3 \epsilon} 
    -\frac{C_A \pi^2}{3} - \frac{10 n_f}{9}+\frac{67 C_A}{9}
    \nn \\
    &\;\;\;\;\;\;\;\;  \qquad\qquad
    +\biggl(\frac{404 C_A}{27} - \frac{56 n_f}{27} - 2 C_A \zeta_3 \biggr) \epsilon  + \mathcal{O}(\epsilon^2)
    \biggr]  \nn \\
    &=  \prnth{\frac{\tilde{\alpha}_s}{4\pi}}S_{1}^{(0)} A(\epsilon) \bigg(
    - \frac{C_A\Gamma_\text{cusp}^{(0)}}{\epsilon^2}
    + \epsilon\,  \frac{\alpha^{(2)}(\epsilon)}{C_A} \bigg)   \,.
\end{align}
Quite remarkably, we find that the one loop soft constant at ${\cal O}(\epsilon)$ is given entirely by the two loop Regge trajectory $\alpha^{(2)}(\epsilon)$!  In Ref.~\cite{Rothstein:2016bsq} the fact that the one soft loop calculation at ${\cal O}(\epsilon^0)$ involved the two-loop cusp anomalous dimension, $\Gamma_{\rm cusp}^{(1)}$, was noted. The analysis done here extends this result to much more intriguing full $\alpha^{(2)}$ from \eq{gregge}. This result is naively quite surprising since it appears that we can obtain two loop anomalous dimensions by carrying out a one-loop calculation. We will explain the reason why this occurs, and argue that it is actually quite natural, in \Sec{sec:relations}. Note that by construction, this soft region radiative correction is independent of the nature of the projectile, and isolates the constant radiative correction associated with the Reggeized gluon.

For future applications, where these one loop corrections will be iterated, it is useful to split the one loop result into its contributions that came from rapidity divergent contributions, and those that are free of rapidity divergences. We therefore take the calculations in \eqs{Sqeye}{Sgeye} and organize them as
\begin{align}
S_1^{(1)}=  \left( \frac{2\alpha^{(1)}(\epsilon)}{\eta}+f(\epsilon)+\cO(\eta)  \right)  \left( \frac{\nu^2}{-t}\right)^{\frac\eta2} +g(\epsilon) \,,
\end{align}
where
\begin{align}
 f(\epsilon)&=A(\epsilon)  \bigg( 
    - \frac{2C_A}{\epsilon^2} 
    -\frac{C_A \pi^2}{3} 
    - 2 C_A \zeta_3\, \epsilon  + \mathcal{O}(\epsilon^2)
    \bigg) \,,  \\
g(\epsilon)&=A(\epsilon)  \bigg( 
    \frac{(11C_A - 2n_f)}{3 \epsilon} 
     - \frac{10 n_f}{9}+\frac{67 C_A}{9}
    +\left(\frac{404 C_A}{27} - \frac{56 n_f}{27}  \right) \epsilon  + \mathcal{O}(\epsilon^2)
    \bigg)\,.    
\end{align}
Here we see that the rapidity divergent induced contribution $f(\epsilon)$ is of uniform transcendental weight (where the weight of $\zeta_n$ is $n$ and weight of $\epsilon$ is $-1$) and independent of the matter content of the theory, since it arises only from Wilson line diagrams. 
The non-rapidity divergent contribution $g(\epsilon)$ contains the matter dependent contributions, and is not of uniform weight.

\subsubsection{$2\to 2$ Glauber Potential}\label{sec:2->2glauber}

We now consider the one-loop soft corrections to the $2\to2$ Glauber potential. We expand the $2\to 2$ Glauber potential perturbatively as
\begin{align} \label{eq:S2expn}
S_{(2,2)}^{(8_A)} \equiv i\pi\, S_2
 =i \pi\,  S_{1}^{(0)} \biggl[ \left( \frac{\tilde \alpha_s}{4\pi} \right) S_2^{(0)}+ \left( \frac{\tilde \alpha_s}{4\pi} \right)^2 S_{2}^{(1)}+\cdots   \biggr] \,.
\end{align}
Since the presence of a Glauber potential loop always leads to a factor of $i\pi$, it is convenient to make this explicit so that the remaining $S_2^{(k)}$ factors are real. Here the base amplitude is given by
\begin{align}  \label{eq:Gbox}
\left(\fd{2.3cm}{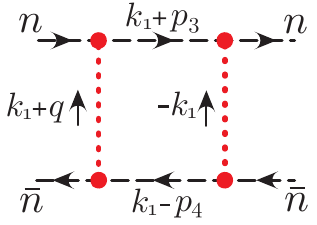}\right)^{(8_A)} 
 &= (i\pi)  \prnth{\frac{\alpha_s}{4\pi}} S_1^{(0)} 
    \biggl( \frac{-t}{4\pi} \biggr)^{\!-\epsilon}\, \frac{B(1,1)}{2}\:  C_A
 \nn\\
 &= (i \pi) \prnth{\frac{\tilde{\alpha}_s}{4\pi}} S_{1}^{(0)}
     \left(-\frac{C_A}{\epsilon}\right) 
\,,
\end{align}
so that we can identify 
\begin{align} \label{eq:Gbox}
  S_2^{(0)} = - \frac{C_A}{\epsilon} = - \frac{\alpha^{(1)}(\epsilon)}{2}  \,.
\end{align} 
Here the subscript emphasizes that this is two Glauber exchange, and the superscript denotes the number of soft loops. We thus see why the modified coupling $\tilde\alpha_s$ is natural -- it arises precisely from the Glauber box integral. The expression for the one-loop Regge trajectory when expanded in $\tilde\alpha_s$ then becomes especially simple.

Note that the existence of the two Glauber exchange graph in \eq{Gbox} also highlights a difference that the Glauber based expansion has from the Reggeon expansion. In the Glauber case it is needed for the effective field theory to be unitary. In the Reggeon case an $8_A$ contribution from an analogous box graph does not exist, and is subsumed by the expansion in objects with definite signature~\cite{Caron-Huot:2013fea}.

\begin{figure}[t!]
\begin{center}
  \raisebox{2cm}{
  \hspace{-1.9cm}
  a)\hspace{3.4cm} 
  b)\hspace{3.8cm} 
  c)\hspace{4.4cm} 
   } \\[-55pt]
\hspace{-0.55cm}
\includegraphics[width=0.19\columnwidth]{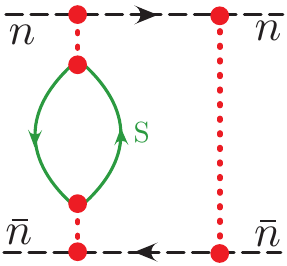}
\hspace{0.6cm}
\includegraphics[width=0.19\columnwidth]{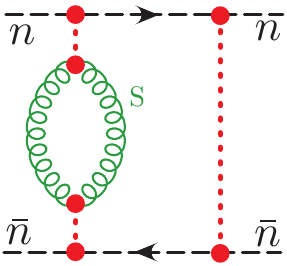}
\hspace{0.4cm}
\includegraphics[width=0.38\columnwidth]{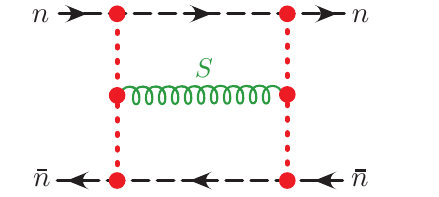}
\\[5pt]
\end{center}
	\vspace{-0.4cm}
	\caption{\setcaptionskip Using crossing symmetry, the two loop Regge trajectory can be computed from three simple graphs. The trivial left-right flipped versions of the first two graphs are present but not shown.}
	\label{fig:diagrams_iter}
	\setmainskip
\end{figure}

The three graphs contributing to the one-loop correction to the $2\to2$ Glauber potential are shown in \Fig{fig:diagrams_iter}. Two of the three graphs are iterations of the one loop graph, and the third is the so called H-graph, which is uniform weight. It is therefore convenient to write
\begin{align}
S_2^{(1)}=S_{2,\text{iterative}}^{(1)}+ S^{(1)}_{2,H}
\end{align}

The results for the iterative graphs follow in a straightforward manner from their one loop counterparts since the extra Glauber loop simply adds a Glauber box integral
\begin{align} \label{eq:Siterative1} 
 2\times & \left(\fd{2cm}{figsSC/Eye_Iteration_Gluon}\right)^{(8_A)}
    + \; 2 \times \left(\fd{2cm}{figsSC/Eye_Iteration_Quark}\right)^{(8_A)} \; 
 \\
 &=i\pi S_1^{(0)} \!\biggl( \frac{\tilde \alpha_s}{4\pi} \biggr)^{\!2}
  C_A
  \biggl\{   2\biggl(\! \frac{2\alpha^{(1)}(\epsilon)}{\eta}\!+\! f(\epsilon)\!+\!\cO(\eta)  \biggr)  
  \biggl( \frac{\nu^2}{-t}\biggr)^{\frac\eta2}
  \frac{B(1,1+\epsilon+\eta/2)}{(-\epsilon) B(1,1)}
 +2 g(\epsilon) \frac{B(1,1+\epsilon)}{(-\epsilon) B(1,1)} \biggr\} .
 \nn
\end{align}
Here the factor of two multiplying each iterated graph accounts for the fact that each soft loop could have appeared on either Glauber. We see that the non-rapidity divergent contributions from $g(\epsilon)$ iterate trivially, since they are just multiplied by the bubble $B(1,1+\epsilon)$. The rapidity dependent terms also iterate, albeit in a slightly more non-trivial manner due to the appearance of $\eta$ in the argument of the Glauber box integral $B(1,1+\epsilon+\eta/2)$. 
Expanding \eq{Siterative1} we can read off the desired constant terms
\begin{align}
    2\times & \left(\fd{2cm}{figsSC/Eye_Iteration_Gluon}\right)^{(8_A)}_{\stackrel{\!\!\!\!\mathcal{O}(\eta^0)}{\nu=\sqrt{-t}} } 
    + \; 2 \times \left(\fd{2cm}{figsSC/Eye_Iteration_Quark}\right)^{(8_A)} \;
 \nn \\
    &=(i \pi) \prnth{\frac{\tilde{\alpha}_s}{4\pi}}^2 S_{1}^{(0)} A(\epsilon) \left( \frac{7}{2\epsilon^3}C_A^2 -\frac{3}{2}\alpha^{(2)}(\epsilon) +\cO(\epsilon)\right)
 \,,
\end{align}    
which corresponds to
\begin{align} \label{eq:S2it}
  S_{2,{\rm iterative}}^{(1)} 
  &= A(\epsilon) \biggl[ \frac{7}{2\epsilon^3} C_A^2 -\frac{3}{2}\alpha^{(2)}(\epsilon) +\cO(\epsilon) \biggr] \,.
\end{align}
The $A(\epsilon)$ is defined in \eq{Aeps} and has an expansion $A(\epsilon) = 1 +{\cal O}(\epsilon^3)$.
We choose to normalize using $A(\epsilon)$ to simplify the structure of \eq{S2it}. In the end when we add various contributions together, the use of $A(\epsilon)$ in the normalization will not have any impact on our final results for the matter dependent contributions, since they only start at $1/\epsilon^2$ in $\alpha^{(2)}(\epsilon)$. 

This result relies strongly on a modified $\eta$ regulator for Glauber operators, which is explained in more detail in \App{sec:regulator}. In \Sec{sec:relations} we will use this result, combined with relations derived from crossing symmetry and unitarity to compute the two-loop Regge trajectory.

For the H-graph, we find
\begin{align}
    \left(\fd{4cm}{figsSC/Hgraph_bare}\right)^{(8_A)}
    &=
    -\frac{\alpha_s^3(\mu) C_A^2 }{4 \, \prpsq{q}} \,
    \frac{\Gma{\frac\eta2} \Gma{\frac{1-\eta}2} 2^{-\eta}}{\sqrt{\pi}} 
    \left(\frac{\mu^2 e^{\gamma_E}}{\prpsq{q}} \right)^{2\epsilon} 
    \left( \frac{\nu^2}{\prpsq{q}}\right)^{\frac\eta2} 
    \nn \\
    &\quad
    \times\bigg[I_1 e^{-2\gamma_E \epsilon}  - 2 B(1,1) \, B\left( 1+\frac\eta2, 1+\epsilon \right)
    \bigg]
   \,.
\end{align}
Here $I_1$ is a two-loop Glauber box integral whose expansion in $\epsilon$ is given by
\begin{align}
    I_1 = \left(\frac4{\epsilon^2} - 4\zeta_2 +\cO(\epsilon) \right) +  \frac\eta2\left(-\frac1{\epsilon^3} + \frac{\zeta_2}\epsilon - \frac{76}3 \zeta_3 +\cO(\epsilon) \right) + \cO(\eta^2) \, .
\end{align}
The integral $I_1$ is known to all orders in $\epsilon$ \cite{Kazakov:1983pk,Kotikov:2018wxe}, although we will not need higher orders in $\epsilon$ for this paper.
The desired constant terms are
\begin{align}
    \left(\fd{4cm}{figsSC/Hgraph_bare}
     \right)^{(8_A)}_{\stackrel{\!\!\!\!\mathcal{O}(\eta^0)}{\nu=\sqrt{-t}} } 
    &=
     (i \pi) \prnth{\frac{\tilde{\alpha}_s}{4\pi}}^2 S_{1}^{(0)} A(\epsilon) C_A^2 
      \left( 
      - \frac3{2 \, \epsilon^3} +\mathcal{O}(\epsilon)
     \right)  \,,
\end{align}
which yields
\begin{align}
  S_{2,H}^{(1)} = A(\epsilon) \biggl[ - \frac{3C_A^2}{2 \, \epsilon^3} +\mathcal{O}(\epsilon) \biggr]\,.
\end{align}
Therefore we see that, as expected, the H-graph is uniform weight, and in fact is a pure $1/\epsilon^3$ pole in our normalization.

Combining the iterated graphs with the H-graph then gives us the $2$-loop, $2$-Glauber soft constant,
\begin{align}
  S_2^{(1)} = 
   S_{2,{\rm iterative}}^{(1)} + S_{2,H}^{(1)} 
   &=  A(\epsilon) \biggl[  \frac{2C_A^2}{\epsilon^3} -\frac{3}{2}\alpha^{(2)}(\epsilon) +\cO(\epsilon) \biggr]\,.
\end{align}

\subsection{Collinear Impact Factors and Glauber Collapse}\label{sec:collinear}

We have seen in the previous section that the soft constants are independent of the collinear projectiles, and purely associated with the dynamics of the Glauber gluons. We therefore expect them to have a direct connection with the Regge trajectory. This has already been made clear by the values of the soft constants, and will be further explained in \Sec{sec:relations}. As discussed earlier, the EFT provides a factorization of the standard impact factors, $C^i = C_s C_c^i$, into the soft factors, $C_s$ and the ``collinear impact factors" $C_c^i$. Here we will argue that 
for the results considered in this work (or more generally under certain approximations), the collinear impact factors are independent of the Glaubers, and depend only on the projectile. Our discussion here will be somewhat brief, including only ingredients necessary for this paper, and leaving a more detailed discussion to Ref.~\cite{collinear_future}.

\begin{figure}[t!]
\begin{center}
  \raisebox{2cm}{
  \hspace{-0.2cm}
  a)\hspace{2.7cm} 
  b)\hspace{2.8cm} 
  c)\hspace{2.7cm} 
  d)\hspace{2.7cm} 
  e)\hspace{3.5cm} 
   } \\[-55pt]
\hspace{-0.55cm}
\includegraphics[width=0.17\columnwidth]{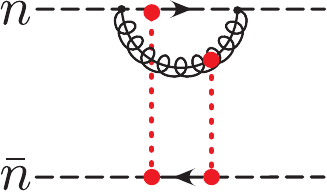}
\hspace{0.2cm}
\includegraphics[width=0.17\columnwidth]{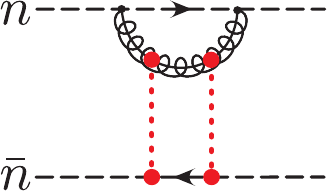}
\hspace{0.2cm}
\includegraphics[width=0.17\columnwidth]{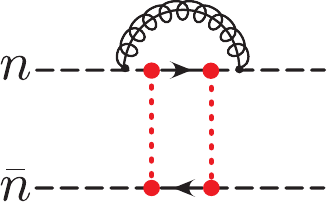}
\hspace{0.2cm}
\includegraphics[width=0.16\columnwidth]{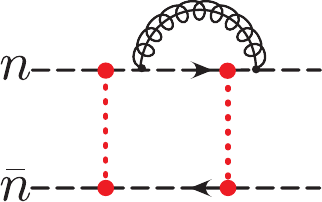}
\hspace{0.2cm}
\includegraphics[width=0.16\columnwidth]{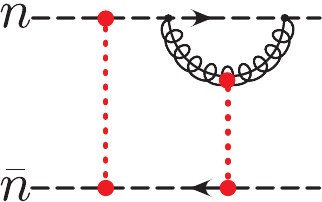}
\hspace{0.2cm}
\\[5pt]
\end{center}
	\vspace{-0.4cm}
	\caption{\setcaptionskip Two-loop mixed Glauber-collinear graphs. The non-planar graph in (a) has a vanishing contribution to the octet channel. The modified $\eta'$ regulator introduced in this paper guarantees an exact factorization to all orders in the regulator of the Glauber and collinear loops in (b), (c) including for the constant terms, as well as enforces the collapse rule to make (d) and (e) vanish.  }
	\label{fig:diagrams}
	\setmainskip
\end{figure}

In this paper we focus on the octet contribution at two-loops, governed by the two-loop gluon Regge trajectory. Because of this, for simplicity, in this section we can restrict ourselves to the planar limit, which suffices for the octet channel at two-loops, as well as in the large $n_f$ limit considered in \Sec{sec:3loop_beyond}. 
This is the case because the non-planar two-loop graph in \fig{diagrams}a does not give a contribution to the $8_A$ color channel.
For planar graphs the collinear constants have a simple interpretation, manifestly independent of the number of Glauber operators.

In \cite{Rothstein:2016bsq} a collapse rule was derived, showing that graphs with multiple Glauber exchanges are non-vanishing only if all Glauber exchanges can collapse towards each other to the same spacetime point, allowing one to reproduce the shockwave picture. In this analysis a common $\eta$ regulator was used for both the soft, collinear and Glauber sectors. At higher loop orders, the picture becomes more intricate, because we encounter diagrams with both Glauber and collinear (or soft) loops, which both require the $\eta$ regulator.  
The Glauber loops never diverge, ie. never give $1/\eta$ factors that are connected to large logarithms, but they do give ${\cal O}(\eta)$ terms, which at two-loops and beyond can interfere with divergent $1/\eta$ contributions from soft and collinear loops.
This results in the presence of $\eta/\eta$ terms that can give finite contributions. For example, in graphs such as  \Fig{fig:diagrams}b,e, we will get an $\eta/\eta$ term that entangles the results from the $1/\eta$-divergent collinear loop and ${\cal O}(\eta)$ terms from the Glauber loop. While the soft and collinear regulators are tied by consistency of the rapidity renormalization group, 
we explain here that the Glauber loops must have a distinct regulator $\eta'$, and that it is important to take $\eta'\to 0$ first in order to correctly reproduce the IR divergences in QCD and also preserve key properties of the EFT at higher loop orders.  
Further details are provided in \App{sec:regulator}.
Thus we have extended the regulator of \cite{Rothstein:2016bsq}, by introducing a distinct regulator $\eta'$ for the Glauber sector, keeping the same $\eta$ regulator for the soft/collinear sectors.  By taking $\eta'\to 0$ prior to expanding in the limit $\eta\to 0$,  terms proportional to $\eta'/\eta \to 0$. 
This modification to the regulator ensures that the collapse rules of SCET with Glauber operators remain true even in the presence of rapidity divergent subdiagrams, yielding a simpler structure of factorization of soft, collinear and Glauber dynamics at higher loop orders. 
In particular, with this modification \fig{diagrams}e vanishes, for the same reason that \fig{diagrams}d vanishes from the collapse rule.

Returning to the example of the two loop mixed collinear and Glauber loop graphs in \Fig{fig:diagrams}, the modified regulator completely factorizes the calculation of the collinear and Glauber dynamics in planar graphs like \Fig{fig:diagrams}b,c.
By first computing the plus and minus Glauber loop integrals, then taking the limit $\eta'\to 0$ with finite $\eta$, we collapse the Glauber exchanges together~\cite{Rothstein:2016bsq}.  This completely eliminates the collinear propagators that are between the two Glauber exchanges, so the double Glauber exchange appears to the remaining parts of the diagram in an identical manner to as if we started with a single Glauber exchange rather than two exchanges.
The result for the remaining collinear loop integral is therefore identical to the one at one lower order, ie. the same for both the single and double Glauber exchange.   For the octet channel this implies the equality of collinear factors given in \eq{J1isJ2}.

This equality for the collinear loop results extends to planar graphs with even more than two Glauber exchanges on the same collinear line. 
This implies that the collinear part of the impact factors, $C_c^i$ are independent of the number of Glaubers in the planar limit, and are a feature only of the projectiles, whereas all dependence on the Glauber dynamics is in the soft factors.  This required modification to the regulator also trivializes the all orders proof of Reggeization in the planar limit, which will be discussed in more detail in \cite{collinear_future}. We believe it is an important step towards an all orders understanding of the structure of the Glauber EFT.
Below in \Sec{sec:relations} we will use unitarity relations to derive the two-loop Regge trajectory. This derivation tests the constants of two loop mixed soft-Glauber and collinear-Glauber diagrams, and provides an additional extremely strong confirmation of the correctness of our modified regulator, as it tests both ${\cal O}(\eta^0)$ and ${\cal O}(\eta'/\eta)\to 0$ terms.

\section{Soft-Glauber Relations and the Two-Loop Regge Trajectory}
\label{sec:relations}

In this section we describe the origin of the iterative structure of the soft function. We will show that this arises from an interesting interplay between crossing symmetry and the modal factorization (soft, collinear, Glauber) in the EFT. This will also make clear why the collinear dynamics plays no role in this structure.

The interplay between crossing symmetry and the Regge limit has a long history, and is encapsulated in the notion of ``signature" in Regge theory. This has played an important role in many recent perturbative studies of the Regge limit. In particular, it implies that high energy logarithms appear in the particular combination $\log(|s|/|t|)- i\pi/2$ \cite{Caron-Huot:2017fxr}.  This simple statement turns out to be  extremely powerful in the EFT approach to the Regge limit due to the fact that logarithms are reproduced by soft or collinear  loops, whereas factors of $i\pi$ are reproduced by Glauber loops. Crossing symmetry therefore has the effect of relating EFT graphs of completely different structures. Most importantly, because Glauber loops are significantly simpler to compute, it allows the exchange of one soft loop for a Glauber loop, effectively dropping the loop order of the calculation.

Although crossing symmetry can be studied directly within the EFT, at the order we work it is conveniently encapsulated into the crossing symmetric expression for the antisymmetric octet exchange for generic projectiles~\cite{Kuraev:1976ge}
\begin{align}
    i \mathcal{M}^{(8_A)}_{2\to2}  \Big|_\text{NLL}
    &= i g_s^2   \left(\frac{s}{t}\right) C_i(p_2, p_3)\left[ \left(\frac{s+i0}{-t}\right)^{\alpha(t)} + \left(\frac{-s-i0}{-t}\right)^{\alpha(t)} \right] C_j(p_1, p_4)\nn \\
    &= i g_s^2  \left(\frac{s}{t}\right) C_i(p_2, p_3) \left(\frac{s}{-t}\right)^{\alpha(t)} \left[ 1 + e^{-i\pi\alpha(t)} \right] C_j(p_1, p_4) \,,
\end{align}
This formula holds to NLL for a general number of colors \cite{DelDuca:2001gu}, and to all orders in the planar limit. Expanding this formula to NLL (i.e. up to two loops in the Regge trajectory), and comparing with our result, we find
\begin{align}
\left( \frac{s}{-t} \right)^{\alpha(t)} J_i J_j [ S_1 + i \pi S_2]= g_s^2 \Big(\frac{s}{t}\Big)\, C_i C_j [2-i\pi \alpha(t) ] \left( \frac{s}{-t} \right)^{\alpha(t)}\,, 
\end{align}
from which we derive that to NLL, we have
\begin{align}
\alpha(t) = -2 \frac{S_2}{S_1}\,.
\end{align}
Expanding this expression perturbatively using Eqs.~(\ref{eq:alphaexpn}, \ref{eq:S1expn}, \ref{eq:S2expn}), we have
\begin{subequations}
\label{eq:alpha12}
\begin{align}
    \alpha^{(1)}(\epsilon) &= -2 S_{2}^{(0)}   \,,
    \label{eq:alpha1}   \\
    \alpha^{(2)}(\epsilon) &=  -2 \Big( S_2^{(1)} -  S_1^{(1)} S_2^{(0)} \Big) \, .
    \label{eq:alpha2}
\end{align}
\end{subequations}
This makes clear that the Regge trajectory can be computed in terms of diagrams with one less soft loop. Beyond NLL, one must incorporate triple Glauber exchange, which we leave to future work. 

Using our perturbative data 
\begin{align} \label{eq:Sdata}
S_2^{(0)}&=-\frac{C_A}{\epsilon} = -\frac{\alpha^{(1)}(\epsilon)}{2}\,,
 \\
S_1^{(1)}&=   A(\epsilon) \bigg(
    - \frac{2C_A}{\epsilon^2}
    + \epsilon\, \frac{\alpha^{(2)}(\epsilon)}{C_A} +\cO(\epsilon^2) \bigg)   \,, 
 \nn \\
S_2^{(1)}&= A(\epsilon) \left( \frac{2}{\epsilon^3}C_A^2 -\frac{3}{2}\alpha^{(2)}(\epsilon) +\cO(\epsilon)\right)
 \,, \nn
\end{align}
we see that the relations in \eq{alpha12} are manifestly true, since we have expressed all our perturbative data in terms of the Regge trajectory. 
For $\alpha^{(2)}(\epsilon)$ the $1/\epsilon^3$ terms cancel between the $S_2^{(1)}$ and $-S_1^{(1)} S_2^{(0)}$ terms.  After this cancellation 
we can use $A(\epsilon)=1+\cO(\epsilon^3)$ and drop the $\cO(\epsilon^3)$ terms, 
which would enter $\alpha^{(2)}(\epsilon)$ only at ${\cal O}(\epsilon)$. 
Thus our results in \eq{Sdata} give
\begin{align}
\alpha^{(2)}(\epsilon) &=\frac{\beta_0}{\epsilon^2}C_A+2 \frac{ \Gamma_\text{cusp}^{(1)}}{\epsilon}C_A  -\frac{56 C_A n_f}{27} + C_A^2 \left( \frac{404}{27} - 2 \zeta_3 \right)\,,
\end{align}
which reproduces the well known result for the two-loop gluon Regge trajectory~\cite{Fadin:1995xg,Fadin:1996tb,Fadin:1995km,Blumlein:1998ib,DelDuca:2001gu}.

\begin{figure}[t!]
	%
	%
	%
	%
	\begin{center}
		\raisebox{2cm}{
			\hspace{-0.4cm}
			a)\hspace{2.8cm} 
			b)\hspace{2.7cm} 
			c)\hspace{2.9cm} 
			d)\hspace{2.7cm} 
			e)\hspace{3.5cm} 
		} \\[-53pt]
		\hspace{-0.5cm}
		\raisebox{0.3cm}{
			\includegraphics[width=0.16\columnwidth]{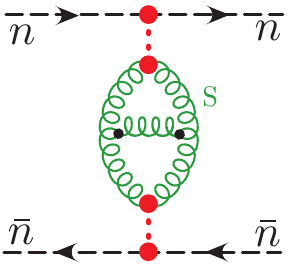}
		}\hspace{0.2cm}
		\raisebox{0.3cm}{
			\includegraphics[width=0.16\columnwidth]{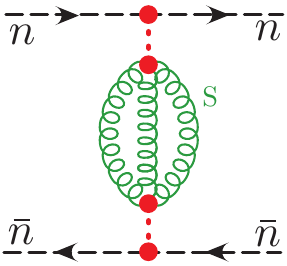} 
		}\hspace{0.2cm}
		\raisebox{0.3cm}{
			\includegraphics[width=0.16\columnwidth]{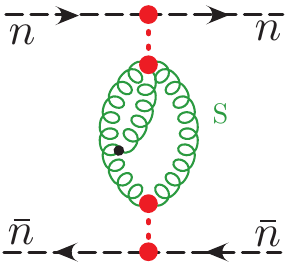}
		}\hspace{0.2cm}
		\raisebox{0.3cm}{
			\includegraphics[width=0.16\columnwidth]{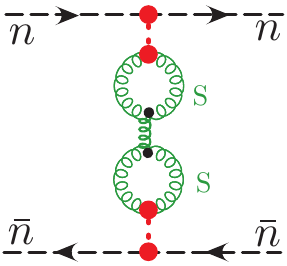}
		}\hspace{0.2cm}
		\raisebox{0.45cm}{
			\includegraphics[width=0.15\columnwidth]{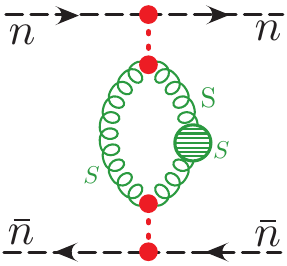}
		} \hspace{0.2cm}
		\\[8pt]
		\raisebox{2cm}{
			\hspace{0.4cm}
			k)\hspace{2.7cm} 
			l)\hspace{2.7cm} 
			m)\hspace{2.1cm} 
			n)\hspace{2.8cm} 
			o)\hspace{3.7cm} 
		} \\[-50pt]
		\hspace{-0.55cm}
		\includegraphics[width=0.15\columnwidth]{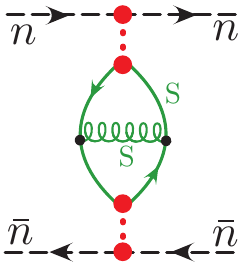}
		\hspace{0.2cm}
		\includegraphics[width=0.15\columnwidth]{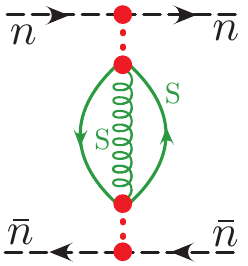}
		\hspace{0.2cm}
		\includegraphics[width=0.15\columnwidth]{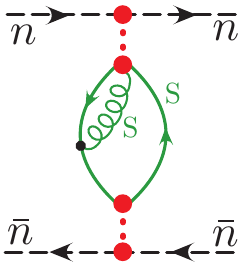}				
		\hspace{0.2cm}
		\includegraphics[width=0.15\columnwidth]{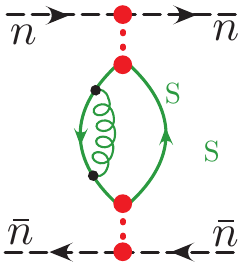}				
		\hspace{0.2cm}
		\includegraphics[width=0.15\columnwidth]{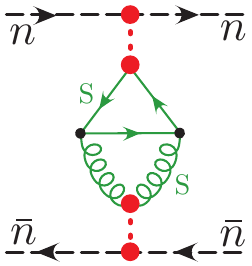}				
	\end{center}
	\vspace{-0.4cm}
	\caption{\setcaptionskip
		Graphs required to directly compute the two-loop Regge trajectory in the EFT (Graphs obtained by left-right or up-down reflections are not shown). Using unitarity and crossing symmetry, this can be reduced to the three graphs shown in \Fig{fig:diagrams_iter}.}
	\label{fig:2loop_SG_graphs}
	\setmainskip
\end{figure}

We believe that our EFT calculation of the two-loop Regge trajectory is interesting for a number of reasons. First, it provides an extremely simple calculation of the two-loop Regge trajectory, involving only the calculation of the three graphs shown in \Fig{fig:diagrams_iter}. This should be compared with the naive soft loop graphs that would be required to calculate the two-loop Regge trajectory in the EFT, which are shown in \Fig{fig:2loop_SG_graphs}.
The classic calculations of the Regge trajectory are also based on two-loop graphs from this soft region of phase space. 
Essentially crossing symmetry allows us to relate one soft loop to a Glauber loop, significantly simplifying the calculation, both in terms of the number, and complexity, of the diagrams involved. 
This also provides a strong check on the structure of the EFT at the two-loop level. In particular, our derivation provides a strong test that the rapidity regulators (see \App{sec:regulator}) in the EFT preserve unitarity.
In \Sec{sec:3loop_beyond}, we will illustrate that this simplification persists at higher perturbative orders by deriving the leading matter contributions to the Regge trajectory at three and four loops. Furthermore, our approach should also allow simple calculations of the Regge trajectory for the quark using the quark Glauber operators derived in  \cite{Moult:2017xpp}.

Secondly, this expression for the Regge trajectory also sheds significant light on the observation of Del Duca \cite{DelDuca:2017pmn} that the two-loop Regge trajectory appears in the one-loop impact factors.  First, we are able to refine his observation. Del Duca observed that the two-loop Regge trajectory could be found in the quark or gluon impact factors although in both cases there was contamination, and in particular, there was significant additional contamination for the case of quarks. In our approach, the two-loop Regge trajectory is found in the constants of the soft function, which are manifestly independent of the scattered projectiles, and so there is no contamination. Due to the unitarity relations and the simple structure of the Glauber graphs in the EFT, we can begin to understand why such an iterative structure is plausible, and why it is in the soft and not the collinear sector. 

First, in \Sec{sec:EFT}, we emphasized that in the EFT the impact factors are further factorized into a soft constant associated with the Glauber potential, and a collinear constant associated with the projectiles. 
It is clear that this factorization is convenient from the perspective of universality; namely, the soft constant is independent of the projectiles. We also believe that there is an important physical distinction, associated with the way that the soft and collinear corrections iterate into higher Glauber graphs. It would be interesting to explore whether it also provides simplifications in theories like ${\cal N}=4$ super Yang-Mills.
In the unitarity relation used above where we have related the two-loop Regge trajectory to the soft constant for two-Glauber exchange, the collinear corrections did not enter, since they are independent of the number of Glaubers, and so factor out. On the other hand, the soft constants feed into higher Glauber loops, leading to an iterative relation. Indeed, in the absence of the H-graph, the relation
\begin{align}
\alpha^{(2)}(\epsilon) &= -2\left(S_{2}^{(1)} - S_{1}^{(1)} S_{2}^{(0)}  \right) \,,
\end{align}
would immediately give an iterative relation between the two-loop Regge trajectory, and the one-loop soft constants. In particular, since the H-graph is matter independent, it immediately proves the relation of Del Duca for the matter dependent terms. In \Sec{sec:3loop_beyond} we will extend this to the leading matter terms at all loops. Furthermore, since the H-graph is maximal weight, it also explains why Del Duca's relation holds for the non-maximal weight terms. The intriguing observation is that the only role of the H-graph is to remove the $1/\epsilon^3$ pole that is present in the iterated graphs, so that it does not appear in the two-loop Regge trajectory. We are unable to prove why this must be the case, however, we can see why there is an iterative structure up to the H-graph. We believe that it would be interesting to explore this at higher loops in the soft sector of the EFT, to understand whether the constants continue to be related to soft anomalous dimensions. We believe this is intriguing, since our arguments constrain the non-maximal weight terms, while the maximal weight terms are constrained to have an iterative form \cite{DelDuca:2008jg} due to the BDS ansatz \cite{Bern:2005iz}.

\section{Leading Matter Dependence at Higher Loops}
\label{sec:3loop_beyond}

The basic argument leading to the relation between soft and Glauber loops is due to unitarity and crossing symmetry, which continue to hold at arbitrary loop order. It is therefore interesting to explore the consequences of our relation beyond two loop order. However, at this order, one encounters the issue that one has both contributions from single Glauber and triple (or higher) Glauber exchange. The definition of the Regge trajectory at this order requires disentangling these contributions, for which there has recently seen significant progress \cite{Caron-Huot:2016tzz,Caron-Huot:2017fxr,Falcioni:2020lvv,Falcioni:2021dgr}.  While this has a natural resolution in the EFT, where the single and triple Glauber exchanges are described by separate gauge invariant operators, it is beyond the scope of the current discussion, and will be left for a future publication.

\begin{figure}[t!]
\begin{center}
\includegraphics[width=0.3\columnwidth]{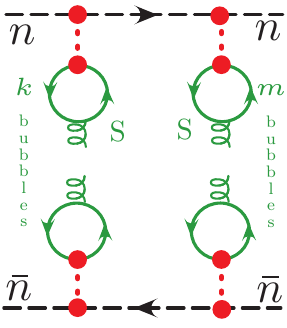}
\end{center}
	\vspace{-0.4cm}
	\caption{\setcaptionskip Graphs needed for the calculation of the leading $n_f$ dependence of the Regge trajectory to all orders in $\alpha_s$.}
	\label{fig:diagrams_nf}
	\setmainskip
\end{figure}

To illustrate the iterative structure at higher loops, we instead choose to focus on the simplest possible example of deriving the leading matter dependence. This has the advantage that one does not have to consider triple (or higher) Glauber exchange, since such contributions have additional gluons, and therefore do not have maximal matter dependence. Further, it implies that we do not have to consider the H-graph, and therefore we have a perfect iterative structure. For these terms we find that the relation
\begin{align}
\alpha(t)= -2 \frac{S_2}{S_1}\,,
\end{align}
holds to all loop orders. This result allows us to derive the maximal matter dependence of the Regge trajectory at any loop order from the one-loop result. It is crucial to emphasize that this result only holds for all terms associated to the \emph{leading} matter dependence, due to the fact that only the class of EFT diagrams in \Fig{fig:diagrams_nf} contribute. For all other color structures, diagrams with additional Glauber gluons, or with gluons connecting the different Glaubers in \Fig{fig:diagrams_nf} (generalized H-graphs), would contribute, and break this relation. By restricting to the leading matter contribution, we are able to extend the use of the relation between one- and two-loop graphs, to an all orders relation. While this is of course the simplest piece of the Regge trajectory, we still find the simplicity with which we are able to derive it quite remarkable.

We can actually write down an explicit description of the maximal matter-dependent pieces of $S_2$ and $S_1$ to all loop orders. For $S_1$, we notice that $S_1$ can be expressed iteratively in terms of the maximal-$n_f$ one-loop graph. We recall that

\begin{align}
    \left(\fd{2cm}{figsSC/Glaub_softvac_quark_ext}\right)^{(8_A)} &= 
  -2 n_f T_F S_1^{(0)}  \frac{\alpha_s(\mu)}{\pi} 
    \left(\frac{\mu^2 e^{\gamma_E}}{-t} \right)^\epsilon 
    \frac{\Gma{2-\epsilon}^2}{\Gma{4-2\epsilon}}  \Gma{\epsilon} 
    \nn \\
    &\equiv S_{1}^{(0)} \mathcal{A}(t) , 
\end{align}
where we define 
\begin{align}
  \mathcal{A}(t) = -2 n_f T_F \frac{\alpha_s(\mu)}{\pi}
    \left(\frac{\mu^2 e^{\gamma_E}}{-t} \right)^\epsilon
    \frac{\Gma{2-\epsilon}^2}{\Gma{4-2\epsilon}}  \Gma{\epsilon} = -\frac{n_f T_f}{\epsilon} \frac{\tilde \alpha_s}{\pi} \frac{1-\epsilon}{(3-2\epsilon)(1 - 2 \epsilon)}. 
\end{align}
Then the sum of the $n$-loop maximal-$n_f$ one-Glauber graphs, obtained by replacing the soft quark bubble by a string of soft bubbles connected by soft gluons, is given by 
\begin{align}
S_1 \Big|_{\alpha_s^{k+1}n_f^k}
 = \frac{S_1^{(0)}}{1 - \mathcal{A}(t) } = \frac{8\pi i \alpha_s}{t} \frac{1}{ 1 + \frac{n_f T_f}{\epsilon} \frac{\tilde \alpha_s}{\pi} \frac{1-\epsilon}{(3-2\epsilon)(1 - 2 \epsilon) }}. 
\end{align}

We can similarly calculate the leading $n_f$ piece of the two-Glauber exchange by simply adding an arbitrary number of soft quark bubbles in between each of the Glauber attachments as shown in \fig{diagrams_nf}.  
Note that a direct calculation of this contribution from purely soft loops would require dressing the $n_f$ dependent graphs in \fig{2loop_SG_graphs} by further quark bubbles, which is a significantly more difficult analysis.
Our result again depends on $\mathcal{A}(t)$:
\begin{align}
S_2\, \Big|_{\alpha_s^{k+2}n_f^k} =
  S_1^{(0)} \frac{\alpha_s C_A}{8\pi}
\Bigl( \frac{4\pi}{-t} \Bigr)^{\epsilon} \!
 \sum_{k, m \geq 0} \!\!
 \frac{ \Gamma(-(k+1)\epsilon) \Gamma(-(m+1) \epsilon) \Gamma(1 +(k+m+1)\epsilon) }{\Gamma(1+k\epsilon) \Gamma(1+m\epsilon) \Gamma( -(k+m+2) \epsilon) } \mathcal{A}(t)^{k+m}\,.
\end{align}
For the terms contributing to the most IR divergent terms, proportional to $\alpha_s^{k+2}n_f^k/\epsilon^{k+1}$ for any $k$,
this sum can be simplified to
\begin{align}
S_2\, \Big|_{\alpha_s^{k+2}n_f^k/\epsilon^{k+1}} = 
  S_1^{(0)} \frac{\alpha_s C_A}{8\pi}
\left( \frac{ 4\pi }{-t} \right)^\epsilon  \frac{1}{\epsilon}  \frac{2 \log\big[1 - \mathcal{A}(t)\big]}{ \mathcal{A}(t)(1 -  \mathcal{A}(t))}
 \,.
\end{align}
Using these results we can compute the maximal $n_f$-dependent piece of the Regge trajectory:
\begin{align} \label{eq:alphanfresult}
\alpha(t)\Big|_{\alpha_s^{k+1} n_f^k} 
&= -2\: \frac{S_2\big|_{\alpha_s^{k+2} n_f^k}  }{S_1\big|_{\alpha_s^{k+1} n_f^k}}
 \\
= -\frac{\alpha_s C_A}{4\pi}& \Bigl( \frac{4\pi}{-t} \Bigr)^{\epsilon} 
\big[ 1 - {\cal A}(t) \big] \!
\sum_{k,m \geq 0} \!
 \frac{ \Gamma\bigl(-(k+1)\epsilon \bigr) \Gamma\bigl(-(m+1) \epsilon\bigr) \Gamma\bigl(1 +(k+m+1)\epsilon\bigr) }
 {\Gamma\bigl(1+k\epsilon) \Gamma\bigl(1+m\epsilon) \Gamma\bigl( -(k+m+2) \epsilon\bigr) } \mathcal{A}(t)^{k+m} 
 \nn \\
= -\frac{\tilde \alpha_s C_A}{4\pi} &
\frac{\big[ 1 - {\cal A}(t) \big]}{(-\epsilon/2) B(1,1)} 
\sum_{k,m \geq 0} \!
 \frac{ \Gamma\bigl(-(k+1)\epsilon \bigr) \Gamma\bigl(-(m+1) \epsilon\bigr) \Gamma\bigl(1 +(k+m+1)\epsilon\bigr) }
 {\Gamma\bigl(1+k\epsilon) \Gamma\bigl(1+m\epsilon) \Gamma\bigl( -(k+m+2) \epsilon\bigr) } \mathcal{A}(t)^{k+m} 
\,, \nn
\end{align}
where $B(1,1)$ is given by \eq{Bab}. 
This is one of our main results, giving the $\alpha_s^{k+1} n_f^k$ terms in Regge trajectory in a form that is valid to all orders in $\epsilon$ and for any $k$. 
For the most singular $1/\epsilon$ poles, the terms proportional to $\alpha_s^{k+1}/\epsilon^{k+1}$ for any $k$, this can be simplified to
\begin{align}
\alpha(t)\, \Big|_{\alpha_s^{k+1}n_f^k/\epsilon^{k+1}} 
 &= 
 -\frac{\alpha_s C_A}{2\pi}
\left( \frac{4\pi}{-t} \right)^\epsilon   \frac{1}{\epsilon}  \frac{ \log\big[ 1 - \mathcal{A}(t) \big]}{ \mathcal{A}(t)} 
 \nn\\
 &= 
-\frac{\tilde \alpha_s C_A}{2\pi}
  \frac{1}{\epsilon}  \frac{ \log\big[ 1 - \mathcal{A}(t) \big]}{ \mathcal{A}(t)} 
\,. 
\end{align}

While our formula in \eq{alphanfresult} allows us to derive the result for the maximal matter dependent terms of the Regge trajectory at any order, here as an application we explicitly give the results for the infrared divergent $1/\epsilon^k$ terms and the ${\cal O}(\epsilon^0)$ terms at order $\alpha_s^3 n_f^2$, $\alpha_s^4 n_f^3$, and $\alpha_s^5 n_f^4$:
\begin{align} \label{eq:alphanfresult345}
\alpha(t) \Big|_{\alpha_s^3 n_f^2} 
&= 
-\frac{\alpha_s C_A}{4\pi}
\left( \frac{4\pi e^{-\gamma_E}}{-t} \right)^\epsilon  \left( \frac{ \tilde \alpha_s n_f T_f }{\pi} \right)^2
 \biggl[ -\frac{2}{27} \frac{1}{\epsilon^3} 
  - \frac{20}{81} \frac{1}{\epsilon^2}
  + \left( \frac{\pi^2}{162} - \frac{2}{3} \right) \frac{1}{\epsilon}
  \nonumber \\ 
&\qquad\quad  + \left( -\frac{1216}{729} + \frac{5\pi^2}{243} + \frac{434 \zeta_3}{81} \right) \biggr]
\nn\\
 &=
-\frac{\tilde \alpha_s C_A}{4\pi}
  \left( \frac{ \tilde \alpha_s n_f T_f }{\pi} \right)^2
 \left[ -\frac{2}{27} \frac{1}{\epsilon^3} 
 - \frac{20}{81} \frac{1}{\epsilon^2} 
  - \frac{2}{3}  \frac{1}{\epsilon} 
 + \left( \frac{140 \zeta_3}{27} -\frac{1216}{729}  \right) \right]
\,, 
\end{align}
\begin{align}
\alpha(t) \Big|_{\alpha_s^4 n_f^3} 
&=
-\frac{\alpha_s C_A}{4\pi}
\left( \frac{4\pi e^{-\gamma_E}}{-t} \right)^\epsilon \left( \frac{ \tilde \alpha_s n_f T_f }{\pi} \right)^3 
\bigg[ \frac{1}{54} \frac{1}{\epsilon^4} 
 + \frac{5}{54} \frac{1}{\epsilon^3} 
 + \left( -\frac{\pi^2}{648} + \frac{53}{162} \right) \frac{1}{\epsilon^2}
  \nonumber \\ 
& \qquad\quad  
+ \left( -\frac{529\zeta_3}{162} - \frac{5\pi^2}{648} + \frac{1457}{1458}  \right) \frac{1}{\epsilon} 
+ \left( -\frac{4223 \pi^4}{77760} - \frac{2645\zeta_3 }{162} - \frac{53\pi^2}{1944} + \frac{2050}{729} \right) \bigg]
 \nn\\
&=
-\frac{\tilde \alpha_s C_A}{4\pi}
 \left( \frac{ \tilde \alpha_s n_f T_f }{\pi} \right)^3 
\bigg[ \frac{1}{54} \frac{1}{\epsilon^4} 
 + \frac{5}{54} \frac{1}{\epsilon^3} 
 + \frac{53}{162} \frac{1}{\epsilon^2} 
 + \left( -\frac{29\zeta_3}{9}  + \frac{1457}{1458}  \right) \frac{1}{\epsilon} 
  \nonumber \\ 
& \qquad\quad
 + \left( -\frac{29 \pi^4}{540} - \frac{145\zeta_3 }{9} + \frac{2050}{729} \right) \bigg]
\,, 
\end{align}
\begin{align}
\alpha(t) \Big|_{\alpha_s^5 n_f^4} 
  &= 
-\frac{\alpha_s C_A}{4\pi}
\left( \frac{4\pi e^{-\gamma_E}}{-t} \right)^\epsilon \left( \frac{ \tilde \alpha_s n_f T_f }{\pi} \right)^4 
\bigg[ -\frac{2}{405} \frac{1}{\epsilon^5} - \frac{8}{243} \frac{1}{\epsilon^4} + \left( \frac{\pi^2}{2430} - \frac{524}{3645} \right) \frac{1}{\epsilon^3}
  \nonumber \\ 
 & \qquad\quad
 + \left( \frac{2102\zeta_3}{1215}+ \frac{2\pi^2}{729} - \frac{5672}{10935} \right) \frac{1}{\epsilon^2} 
+ \left( \frac{8399\pi^4}{291600} + \frac{8408\zeta_3}{729} + \frac{131\pi^2}{10935} - \frac{54946}{32805} \right) \frac{1}{\epsilon} 
  \nonumber \\
 & \qquad\quad + \left( \frac{4426\zeta_5 }{75}  + \frac{8399\pi^4}{43470} + \frac{550724\zeta_3}{10935} - \frac{1051 \pi^2 \zeta_3}{7290} + \frac{1418\pi^2}{32805} - \frac{494528}{98415}\right)
\bigg]
\nn\\
&=
-\frac{\tilde\alpha_s C_A}{4\pi}
 \left( \frac{ \tilde \alpha_s n_f T_f }{\pi} \right)^4 
\bigg[ -\frac{2}{405} \frac{1}{\epsilon^5} 
 - \frac{8}{243} \frac{1}{\epsilon^4} 
 - \frac{524}{3645}  \frac{1}{\epsilon^3}
 + \left( \frac{232\zeta_3}{135} - \frac{5672}{10935} \right) \frac{1}{\epsilon^2} 
  \nonumber \\ 
 & \qquad\quad
+ \left( \frac{58\pi^4}{2025} + \frac{928\zeta_3}{81} - \frac{54946}{32805} \right) \frac{1}{\epsilon} 
+ \left( \frac{23888\zeta_5 }{405}  
 + \frac{232\pi^4}{1215} + \frac{60784\zeta_3}{1215}  - \frac{494528}{98415}\right)
\bigg]
 \,.
\end{align}

There are a number of cross checks on these results. First, our result for the leading-$n_f$ piece of the three-loop Regge trajectory $\alpha|^{\alpha_s^3n_f^2}_{n_f}$, including the $\cO(\epsilon^0)$ terms, agrees with the recent explicit calculations \cite{Caola:2021rqz,Falcioni:2021dgr} (see also \cite{DelDuca:2021vjq} for an earlier calculation of the planar pure Yang-Mills contributions). Note that our conventions differ from those used in \cite{Caola:2021rqz}; their three-loop Regge trajectory includes the order-$\alpha_s^3$ contributions from the running coupling $\alpha_s$, while our expression is written in terms of the bare coupling $\alpha_s$ or $\tilde\alpha_s$. In addition, the singular terms (containing $1/\epsilon$ or higher poles) in our expressions for the leading-$n_f$ Regge trajectory $\alpha|^{\alpha_s^4n_f^3}_{n_f}$ and $\alpha|^{\alpha_s^5n_f^4}_{n_f}$ agree at four- and five-loop order with the prediction from infrared factorization \cite{Korchemskaya:1994qp,Korchemskaya:1996je,Bret:2011xm,DelDuca:2011ae,DelDuca:2013ara,Falcioni:2021buo}:
\begin{align}
K(\alpha_s) = -\frac{1}{4} \int_0^{\mu^2} \frac{d\lambda^2}{\lambda^2} \hat{\gamma}_K\big(\alpha_s(\lambda)\big). 
\end{align}
Here $K(\alpha_s)$ denotes the $\epsilon$-divergent terms in the Regge trajectory, and the formula above relates $K(\alpha_s)$ to the cusp anomalous dimension $\hat{\gamma}_K$. 
Here and below we use the notation $\hat\gamma_K = \Gamma_{\rm cusp}$ and $\hat\gamma_K^{(i)} = \Gamma_{\rm cusp}^{(i)}$ to facilitate the comparison to the literature~\cite{DelDuca:2011ae,DelDuca:2013ara}.
Explicit expressions for $K(\alpha_s)$ are known to four loops \cite{Korchemsky:1987wg,Moch:2004pa,Henn:2019swt,Huber:2019fxe}. The maximally matter dependent terms of $\hat\gamma_K$ are also known \cite{Gracia:2021}. Writing 
\begin{align}
  \hat{\gamma}_K &= \sum_{i \geq 1} \hat{\gamma}_K^{(i)} \left( \frac{\alpha_s(\lambda)}{\pi} \right)^i 
   = \sum_{i \geq 1} \hat{\gamma}_K^{(i)} \left[
 \frac{Z_\alpha(\alpha_s(\mu),\epsilon)}{Z_\alpha(\alpha_s(\lambda),\epsilon)} \Big(  \frac{\mu^2}{\lambda^2}\Big)^\epsilon 
 \:\frac{\alpha_s(\mu)}{\pi} \right]^i
  \,,\\
 K(\alpha_s) &= \sum_{i \geq 1} K^{(i)} \left( \frac{\alpha_s}{\pi} \right)^i
  \,,\nn
\end{align}
we find the following expressions for $K^{(4)}$ and $K^{(5)}$:
\begin{align}
K^{(4)} &= -\frac{\beta_0 \hat{\gamma}_K^{(1)}}{1024 \epsilon^4} + \frac{\beta_0\beta_1 \hat{\gamma}_K^{(1)}}{128\epsilon^3} + \frac{\beta_0^2 \hat{\gamma}_K^{(2)}}{256 \epsilon^3} - \frac{\beta_2 \gamma_K^{(1)}}{64\epsilon^2} - \frac{\beta_1 \hat{\gamma}_K^{(2)}}{64\epsilon^2} - \frac{\beta_0 \hat{\gamma}_K^{(3)}}{64\epsilon^2} + \frac{\hat{\gamma}_K^{(4)}}{16\epsilon}
 \,, \\
K^{(5)} &= \frac{\beta_0^4 \hat{\gamma}_K^{(1)}}{5120\epsilon^5} - \frac{3\beta_0^2 \beta_1 \hat{\gamma}_K^{(1)}}{1280\epsilon^4} - \frac{\beta_0^3 \hat{\gamma}_K^{(2)}}{1280\epsilon^4} + \frac{\beta_1^2 \hat{\gamma}_K^{(2)}}{320 \epsilon^2} + \frac{\beta_0 \beta_2 \hat{\gamma}_K^{(1)}}{160\epsilon^3} + \frac{\beta_0\beta_1 \hat{\gamma}_K^{(2)}}{160\epsilon^3} 
 \nn \\
&\quad + \frac{\beta_0^2 \hat{\gamma}_K^{(3)}}{320\epsilon^3} - \frac{\beta_3 \hat{\gamma}_K^{(1)}}{80\epsilon^2} - \frac{\beta_2 \hat{\gamma}_K^{(1)}}{80 \epsilon^2} - \frac{\beta_1 \hat{\gamma}_K^{(3)}}{80 \epsilon^2} - \frac{\beta_0 \hat{\gamma}_K^{(4)}}{80\epsilon^2} + \frac{\hat{\gamma}_K^{(5)}}{20\epsilon}\,.
 \nn
\end{align}
Using the all-orders leading-$n_f$ cusp anomalous dimension in \cite{Gracia:2021}, which we re-express in terms of the bare coupling, we find perfect agreement between the $K^{(i)}$ and the singular pieces, $1/\epsilon^k$ for $k\ge 1$, of our leading-$n_f$ Regge trajectory up to five-loop order in \eq{alphanfresult345}. This provides further evidence in support of the conjecture in \cite{Falcioni:2021buo}. 

To the best of our knowledge, the constant $\cO(\epsilon^0)$ terms of the leading-$n_f$ Regge trajectory at four- and five-loop order in \eq{alphanfresult345} are new.  Our more general result in \eq{alphanfresult} can be used to make analogous predictions for leading $n_f$ terms at higher orders in $\alpha_s$, as well as higher orders in $\epsilon$.

\section{Conclusions}
\label{sec:conc}

In this paper we have reconsidered the $2\to2$ scattering amplitude in the Regge limit from the perspective of the Glauber EFT \cite{Rothstein:2016bsq}. While various forms of factorization in the high energy limit have of course long been appreciated, in this paper we have argued that the standard factorization into impact factors, obscures some of the underlying simplicity of the constants appearing in these functions.

Using the Glauber EFT, we factorized the amplitude into separate gauge invariant soft and collinear functions. The soft functions are universal (i.e. independent of the projectiles) and describe radiative corrections to the Reggeized gluons. We computed these universal functions at one loop to $\cO(\epsilon)$ in dimensional regularization. We found, quite remarkably, that they are expressed in terms of two loop anomalous dimensions of Wilson line configurations, without any contamination. 

We then argued that this iterative structure follows from the action of crossing symmetry on graphs in the EFT. Since the EFT factorizes loops into soft, collinear and Glauber modes, crossing symmetry acts non-trivially and relates contributions from different modes. In particular, we find that it can be used to eliminate a soft loop in exchange for a much simpler Glauber loop. Using this, we were able to provide a simple calculation of the two-loop Regge trajectory, and also explain some of the iterative structures observed in \cite{DelDuca:2017pmn}.

We also explored the structure of the iterations at higher orders. In particular, as a simple application, we were able to derive the maximal-$n_f$ terms in the Regge trajectory to all orders in $\alpha_s$.  We checked our result against the recent $\alpha_s^3$ calculation of \cite{Falcioni:2021dgr,Caola:2021izf,DelDuca:2021vjq} finding perfect agreement. We also provided explicit results for the $\alpha_s^4 n_f^3$ and $\alpha_s^5 n_f^4$ terms in the Regge trajectory. We find it quite interesting that this relation allows us to show iterative properties of the maximal-$n_f$ terms, while iteration of the maximally transcendental pieces follows from their relation to $\cN=4$ SYM. We hope that further understanding may enable some iterative properties of QCD amplitudes to emerge in the Regge limit.

There are many directions for future study. Using an identical approach combined with the Glauber quark operators \cite{Moult:2017xpp} (see \cite{Bhattacharya:2021zsg} for a verification of these operators at one-loop order), one should be able to give a simple direct calculation of the two-loop quark Regge trajectory in QCD/QED, which has so far only be derived from direct expansion of the two loop amplitudes \cite{Bogdan:2002sr}. Also at the two-loop level, it would be interesting to understand possible iterative structures in other color channels, for example the pomeron.

More non-trivially, it will be extremely interesting to understand the structure of the iterative relations at the three-loop level. This requires an understanding of the interplay between one- and three-Glauber exchange in the EFT, but we are hopeful that the clean factorization of the EFT may also allow interesting patterns to be unravelled at this order. There has recently been significant progress in understanding the structure of multiple Reggeon exchanges from a variety of different approaches \cite{Caron-Huot:2017fxr,Falcioni:2021dgr,Falcioni:2020lvv}, which resulted in the direct determination of \cite{Caola:2021izf,DelDuca:2021vjq,Falcioni:2021dgr}. An independent calculation of this result using the EFT would be quite interesting.

We are optimistic that the organization of the EFT for forward scattering can provide new insights into the Regge limit, and  that despite many years of intensive study there are still interesting surprises hinting at much more structure yet to be discovered. 

\begin{acknowledgments}
We thanks Vittorio Del Duca, Hua Xing Zhu, Duff Neill, George Sterman, Lance Dixon for interesting discussions and for sharing their numerological  observations. We also thank Anjie Gao for providing many helpful comments on the draft.
This work was supported by the U.S.~Department of Energy, Office of Science, Office of Nuclear Physics, from DE-SC0011090. I.M. was supported by start-up funds from Yale University. I.S.~was also supported in part by the Simons Foundation through the Investigator grant 327942. 

\end{acknowledgments}

\appendix  

\section{Rapidity Regulators for Glauber Potentials and Soft Operators}\label{sec:regulator}

For calculations in SCET with Glauber operators, rapidity regulators are required, both for Glauber loops which lead to $i\pi$ factors, and for soft and collinear loops where they are associated with the rapidity logarithms. We follow the regulators of Ref.~\cite{Rothstein:2016bsq} for this purpose, which are based on the rapidity regulator $\eta$ of Ref.~\cite{Chiu:2012ir}. However, we have found that some modifications and additions are needed to fully and consistently regulate diagrams at two-loop order and beyond in the Glauber EFT.  Those changes are summarized here. 

The first change we make is to decouple the regulator used for Glauber loops from that used for soft and collinear loops.  Usually in an EFT it is important to consistently use the same regulator for the EFT loops, since the regulator can act to separate contributions in different regions, and must do so consistently.  Indeed this is the case for rapidity divergences in soft and collinear loops that are associated with Reggeization, where these loops individually have $1/\eta$ divergences that cancel in their sum, see~\cite{Rothstein:2016bsq}. 
We observe however that the regulator for Glauber loops acts in a different manner, it leads to finite results as $\eta\to 0$, and these results are exactly the terms necessary to preserve unitarity for forward scattering in the EFT.  Therefore it is reasonable that a different regulators $\eta'$ can be used for the Glauber loops.  
However we argue that it is not only reasonable, but necessary to the consistency of the EFT to separate these regulators, and to take $\eta'\to 0$ prior to taking $\eta\to 0$.  To see this one can examine the diagram in \fig{diagrams}e.  The collinear sub-loop in this diagram has a $1/\eta$ divergence. However, by the arguments of Ref.~\cite{Rothstein:2016bsq} this diagram should have a Glauber loop that causes the diagram to vanish, since the Glauber potentials are interrupted by the collinear gluon vertex from collapsing onto the same longitudinal position, leading to terms $\propto \eta'/\eta$. Consistency of the EFT requires that the collapse rule, at the heart of the $(t,z)$ instantaneity for the Glauber potential, be maintained, and therefore that we take $\eta'\to 0$ prior to expanding about $\eta\to 0$.  Further details of this discussion are left to Ref.~\cite{collinear_future}.  To utilize independent Glauber regulators we write
\begin{align} \label{eq:LGreg_eta}
	\int\!\!  d^4x\, {\cal L}_G^{{\rm II}(0)}  & = 
	\sum_{n,\bn} \sum_{i,j=q,g} \! \int\! [dx^\pm] \!
	\sum_{k_r^+,k_r^-} 
	\int\!\! \frac{d^2q_\perp}{q_\perp^2} \frac{d^2q_\perp^\prime}{q_\perp^{\prime\,2}} \: 
	{\cal O}_{s,-k_r^\pm}^{AB}(q_\perp,q_\perp')
	\\
	&\qquad\qquad \times
	\Bigg[ {\cal O}_{n,k_r^-}^{i A}(q_\perp)\,  
	w ^{\prime\,2} \Bigg|\frac{in\,\cdot\! \leftpartial
		+i\bn\,\cdot\! \vec\partial
	}{\nu'} \Bigg|^{-\eta'}
	\, {\cal O}_{\bn,k_r^+}^{jB}(-q_\perp^\prime)
	\bigg] 
	\nn\\
	&\hspace{-1.4cm}
 + \sum_n \sum_{i,j=q,g}  \int\! [dx^\pm] \,
	\sum_{k_r^-} \!\!   \int\!\! \frac{d^2q_\perp}{q_\perp^2} \: 
	{\cal O}_{n,-k_r^-}^{i A}(q_\perp)\,  w' \Bigg|\frac{-\beta_{ns}\,k_r^- - i\bn\cdot \leftpartial - i n \cdot \vec{\partial} }{\nu'}\Bigg|^{-\eta'/2} 
	\, {\cal O}_{s,k_r^-}^{j_n A}(-q_\perp)  
	\,.  \nn
\end{align}
Here the rapidity scale $\nu'$ and book-keeping parameter $w'$ (whose renormalized value is $1$) will be irrelevant to the result for Glauber loop graphs, which are all finite as $\eta'\to 0$.  In addition we have used a 
regulator for the soft-collinear part of the Glauber potential with a power $-\eta'/2$.  This ensures that graphs like those in \fig{diagrams_iter}a,b involve the same regulating factor for their Glauber loops as the base box diagram in \eq{Gbox}.  While this does not have a direct impact on the calculations carried out here, it does ensure that the exponentiation associated to the eikonal phase works out properly in the presence of soft fluctuations of the Glauber potentials. 

Next we discuss an extension of the regulating factors $\eta$ for soft and collinear Wilson lines used in~\cite{Rothstein:2016bsq}, which we find are necessary beyond one-loop order. At two loop order there are rapidity divergences in diagrams that do not involve Wilson line Feynman rules in our  operator basis. Also certain cancellations associated to the equations of motion can be blocked when regulators only appear in Wilson lines. This requires additional regulating factors for both collinear and soft operators. In particular, we find that it is important  for the overall regulator to appear homogeneously in the one-gluon $B_{s\perp}^{n\mu}$ Feynman rule, and in the soft gluon regulator for the Lipatov vertex.  Again the detailed arguments for this are left to Ref.~\cite{collinear_future}, and we simply summarize the modified regulators here.  A list of operators with additional regulators is given in Tab.~\ref{table:opsummary}. 
\begin{table}[t]
	\fontsize{10}{7}\selectfont
	\line(1,0){470}
	\vspace{-0.2cm}
	\begin{align}
		{\cal O}_n^{g B} & = \frac{i}{2} f^{BCD}  {\cal B}_{n\perp\mu}^C \,
		\frac{\bn}{2}\cdot (\cP\!+\!\cP^\dagger)
		w^{-2} \left|\frac{\bn}{2} \cdot \frac{(\cP\!+\!\cP^\dagger)}{\nu}\right|^{\eta}  
		{\cal B}_{n\perp}^{D\mu} 
		\nn\\
		{\cal O}_s^{\! BC} & ={8\pi\alpha_s  }
		\bigg\{
		\cP_\perp^\mu {\cal S}_n^T {\cal S}_\bn  \cP_{\perp\mu}
		- \cP^\perp_\mu g \widetilde {\cal B}_{S\perp}^{n\mu}  {\cal S}_n^T  {\cal S}_\bn 
		-  {\cal S}_n^T  {\cal S}_\bn  g \widetilde {\cal B}_{S\perp}^{\bn\mu} \cP^\perp_{\mu} 
		\nn\\ 
		&  \qquad\qquad\;\; -  g \widetilde {\cal B}_{S\perp}^{n\mu}  {\cal S}_n^T  {\cal S}_\bn g \widetilde {\cal B}_{S\perp\mu}^{\bn}
		-w \left|\frac{2 \cP^z}{\nu}\right|^{-\eta/2} \frac{n_\mu \bn_\nu}{2} {\cal S}_n^T   ig \widetilde {G}_s^{\mu\nu} {\cal S}_\bn 
		\bigg\}^{BC} 
		\nn\\
		{\cal O}_s^{g_n B}  &= 8\pi\alpha_s\: \Big(
		\frac{i}{2} f^{BCD}  {\cal B}_{S\perp\mu}^{n C}\, 
		\frac{n}{2} \cdot (\cP\!+\!\cP^\dagger)  
		w \left| \frac{2\cP_z\!+\!2\cP_z^\dagger}{\nu}\right|^{\eta/2}  
		{\cal B}_{S\perp}^{n D\mu} \Big)  
		\nn\\
       	W_n &= \sum_{\rm perms} \exp\bigg\{ \frac{-g}{\bn\cdot \cP} \bigg[
		\frac{w^2 |\bn\cdot \cP|^{-\eta}}{\nu^{-\eta}} \bn \cdot A_n \bigg]\bigg\}
		\,, \qquad
		S_n = \sum_{\rm perms} \exp\bigg\{ \frac{-g}{n\cdot \cP} \bigg[
		w \left|\frac{2 \cP^z}{\nu}\right|^{-\eta/2} n \cdot A_s\bigg] \bigg\} 
        \nn\\
		{\cal B}_{n\perp}^\mu &= \left(A_{n\perp}^\mu w \left|\frac{\bn\cdot \cP}{\nu}\right|^{-\eta/2} - \frac{k_\perp^\mu}{\bn\cdot k} \bn\cdot A_{n,k} w^2 \left|\frac{\bn\cdot \cP}{\nu}\right|^{-\eta} \right) + \ldots  
		\nn\\
		{\cal B}_{s\perp}^{n\mu} &=
        \frac{1}{g} \frac{1}{n\cdot\cP} 
        w \left|\frac{2 \cP^z}{\nu}\right|^{-\eta/2}
        n_\nu i G_s^{B\nu\mu} {\cal S}_n^{BA} T^A
        = w \left|\frac{2 \cP^z}{\nu}\right|^{-\eta/2} 
        \left(A_{s\perp}^\mu - \frac{k_\perp^\mu}{n\cdot k} n\cdot A_{s,k}\right) + \ldots  
		\nn\\[-20pt] \nn
	\end{align}
	\line(1,0){470}
	\vspace{-.1cm}
	\caption{
		\label{table:opsummary}
		Summary of regulators within the operators appearing in the leading power Glauber exchange Lagrangian.
		For all operators except ${\cal O}_s^{\! BC}$ there exists an analagous $ n \leftrightarrow \bn$ operator.
	}
\end{table}
These regulators modify the Feynman rules involving soft gluons listed in Ref.~\cite{Rothstein:2016bsq} 
so that they are now (setting $\eta'=0$ for simplicity)
\begin{align}
%
%
	\fd{3cm}{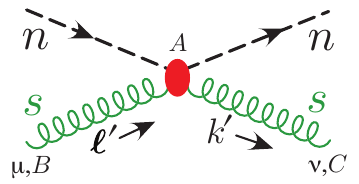} 
  & 
  = \frac{-8\pi \alpha_s f^{ABC} }{(\vec \ell^\prime_\perp-\vec k^\prime_\perp)^2 }
	\Big[ \bar u_n \frac{\bnslash}{2} T^A u_n  \Big]  
	\Big[ n\mcdot k^\prime\, g_\perp^{\mu\nu} - n^\mu \ell^{\prime\nu}_\perp -n^\nu k^{\prime\mu}_\perp 
	+ \frac{\ell^\prime_\perp\cdot k^\prime_\perp n^\mu n^\nu}{n\cdot k^\prime} \Big] 
  \nn\\
 &\quad \times 
  w \left|\frac{n\cdot k' - \bn \cdot l'}{\nu}\right|^{-\eta/2}
   \left|\frac{n\cdot k' - \bn \cdot k'}{\nu}\right|^{-\eta/2}
    \left|\frac{2n\cdot k' - \bn \cdot (k'+\ell')}{\nu}\right|^{+\eta/2}
\,,
\end{align}
where we note that the soft $n\cdot \ell' = n\cdot k'$, and for the Lipatov vertex we have
\begin{align}
%
%
	\fd{3cm}{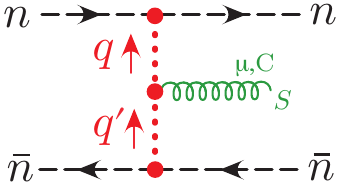} & \!\! 
  = i \Big[ \! \bar u_n \frac{\bnslash}{2} T^A u_n \! \Big] \!
	\biggl[ \frac{8\pi\alpha_s}{\vec q_\perp^{\,2}\, \vec q_\perp^{\,\prime 2} } 
	\,  ig f^{ABC} \biggl( q_\perp^\mu\!+\!q_\perp^{\prime\mu} 
	- n\cdot q^\prime \frac{\bn^\mu}{2}
	-\bn\cdot q \frac{n^\mu}{2}
	- \dfrac{n^\mu \vec q_\perp^{\: 2}}{n\cdot q'} 
	- \dfrac{\bn^\mu \vec q_\perp^{\: \prime 2}}{\bn\cdot q}  
	\biggr) \biggr]  \nn \\
	& \;\; \times \! \Big[ \! \bar v_\bn \frac{\nslash}{2} \bar T^{B} v_\bn\!  \Big] 
   w \left|\frac{\bn\cdot q - n \cdot q'}{\nu}\right|^{-\eta/2} 
  \,.
\end{align}
For completeness, we also list the new Feynman rule for the collinear sector (again setting $\eta'=0$ for simplicity),
\begin{align}
	\fd{3cm}{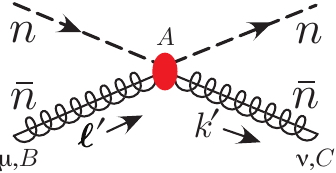} & = \frac{-8\pi \alpha_s f^{ABC} }{(\vec \ell^\prime_\perp-\vec k^\prime_\perp)^2 }
	\Big[ \bar u_n \frac{\bnslash}{2} T^A u_n  \Big]  
	\\*
	& \quad \times \bigg[ n\mcdot k^\prime\, g_\perp^{\mu\nu} 
	- \left(n^\mu \ell^{\prime\nu}_\perp + n^\nu k^{\prime\mu}_\perp \right) w \left| \frac{n \cdot k'}{\nu}\right|^{-\eta/2}
	+ \frac{\ell^\prime_\perp\!\cdot k^\prime_\perp n^\mu n^\nu}{n\cdot k^\prime} w^2 \left| \frac{n \cdot k'}{\nu}\right|^{-\eta} \bigg]   
  .\nn
\end{align}
For $\bn$-collinear gluons scattering with $n$-collinear quarks we have a similar Feynman rule, just with $n\leftrightarrow \bn$. Also for a $n$-collinear gluon scattering with an $\bn$-collinear gluon, the same combinations of regulating factors appear in each of the two-gluon parts of the Feynman rule.

\bibliographystyle{JHEP}
\bibliography{../reggeNLL_bib}

\end{document}